\documentclass[fleqn,usenatbib]{mnras}

\usepackage{newtxtext,newtxmath}
\usepackage{graphics}
\usepackage{epsfig}	
\usepackage{amsmath}	

\usepackage{amssymb}	

\usepackage[T1]{fontenc}

\DeclareRobustCommand{\VAN}[3]{#2}
\let\VANthebibliography\thebibliography
\def\thebibliography{\DeclareRobustCommand{\VAN}[3]{##3}\VANthebibliography}

\usepackage[switch]{lineno}
\usepackage{xspace}

\let\oldequation\equation
\let\oldendequation\endequation

\renewenvironment{equation}{\linenomathNonumbers\oldequation}{\oldendequation\endlinenomath}

\newcommand{\dd}{{\rm d}}
\newcommand{\einn}{\texttt{EinN04}\xspace}
\newcommand{\einw}{\texttt{EinW22}\xspace}

\title[Exploring DM spike distribution around GC with stellar orbits]{Exploring dark matter spike distribution around the Galactic centre with stellar orbits}

\author[Zhao-Qiang Shen et al.]{
Zhao-Qiang Shen$^{1}$,
Guan-Wen Yuan$^{1,2}$,
Cheng-Zi Jiang$^{2,3}$,
Yue-Lin Sming Tsai$^{1}$\thanks{E-mail: smingtsai@pmo.ac.cn (YLST), yzfan@pmo.ac.cn (YZF)},
Qiang Yuan$^{1,2}$,
\newauthor
Yi-Zhong Fan$^{1,2}$\footnotemark[1]
\\
$^{1}$Key Laboratory of dark Matter and Space Astronomy, Purple Mountain Observatory, Chinese Academy of Sciences, Nanjing 210033, China \\
$^{2}$School of Astronomy and Space Sciences, University of Science and Technology of China, Hefei 230026, China\\
$^{3}$Key Laboratory of Planetary Sciences, Purple Mountain Observatory, Chinese Academy of Sciences, Nanjing 210033, China \\
}

\date{Accepted 2023 October 23. Received 2023 October 02; in original form 2023 March 26 }

\pubyear{2023}

\begin{document}
\label{firstpage}
\pagerange{\pageref{firstpage}--\pageref{lastpage}}
\maketitle

\begin{abstract}
Precise measurements of the stellar orbits around Sagittarius A* have established the existence of a supermassive black hole (SMBH) at the Galactic centre (GC).
Due to the interplay between the SMBH and dark matter (DM), the DM density profile in the innermost region of the Galaxy, which is crucial for the DM indirect detection, is still an open question. 
Among the most popular models in the literature, 
the theoretical spike profile proposed by Gondolo and Silk (1999; GS hereafter) is well adopted.
In this work, we investigate the DM spike profile using updated data from the Keck and VLT telescopes considering that the presence of such an extended mass component may affect the orbits of the S-stars in the Galactic centre.
We examine the radius and slope of the generalized NFW spike profile, analyze the Einasto spike, and discuss the influence of DM annihilation on the results.
Our findings indicate that an initial slope of $\gamma \gtrsim 0.92$ for the generalized NFW spike profile is ruled out at a 95\% confidence level.
Additionally, the spike radius $R_{\rm sp}$ larger than 21.5~pc is rejected at 95\% probability for the Einasto spike with $\alpha=0.17$,
which also contradicts the GS spike model.
The constraints with the VLT/GRAVITY upper limits are also projected.
Although the GS NFW spike is well constrained by the Keck and VLT observation of S2, an NFW spike with a weak annihilation cusp may still be viable, as long as the DM annihilation cross section satisfies $\left< \sigma v \right> \gtrsim 7.7\times 10^{-27}~{\rm cm^3\,s^{-1}}  (m_{\rm DM}/100~{\rm GeV})$ at 95\% level.
\end{abstract}


\section{Introduction}
The nature of dark matter (DM), an invisible component that provides the additional gravitational force necessary to explain a range of phenomena across different scales, remains one of the most significant enigmas in the universe~\citep{Bergstrom2000,Bertone2018}. 
To date, the particle nature of DM remains largely unknown, despite numerous proposals for DM candidates in the scientific literature~\citep{Bertone:2004pz,Feng:2010gw,Hu:2000ke, deLaurentis2022}. 
In addition, some systematic searches have been conducted to explore these candidates~\citep{Poter2011,Charles2016,Roszkowski2018,Liu:2017drf,2022PhRvL.129w1101Z}.
Of particular interest in the indirect detection of DM is the Galactic centre (GC), 
where the density of DM peaks and certain excesses have been reported and investigated~\citep[e.g.][]{Hooper2011,DiMauro2021,Cholis2022,Bringmann2012,2015PhRvD..91l3010Z,Ackermann2015,DAMPE:2021hsz,MAGIC2023line}. 
However, it is challenging to determine the DM density in the inner Galaxy based on the rotation curve of interstellar gas, as outlined in previous studies~\citep{Sofue2013}.

The actual DM density profile may theoretically differ from the DM-only halo as a result of the interplay between DM and the super-massive black hole (SMBH)~\citep[e.g.][]{Detweiler1980,Gondolo:1999ef,Yuan2022a, 2023arXiv230109403C}.
The DM spike, a DM structure even steeper than the cusp, can be formed along with the growth of the SMBH.
The accumulation of DM particles in the GC can be comprehensively explained as a result of the inward flow of the ordinary interstellar medium, which undergoes accretion by the black hole (BH) due to frictional forces. 
This process leads to an increase in the mass and gravitational potential of the BH, 
ultimately leading to the accumulation of DM particles in the GC. 
According to the proposal by \citet{Gondolo:1999ef}, if the SMBH in the GC grows adiabatically, 
the DM density could be enhanced by up to ten orders of magnitude.
The profile (GS spike hereafter) is proportional to $r^{-\gamma_{\rm sp}}$, 
with the spike slope range $2.25<\gamma_{\rm sp}<2.5$~\citep{Gondolo:1999ef,Merritt2002a,Sadeghian2013,Ferrer2017}.
This spike profile has been widely adopted in previous studies, 
leading to strong constraints on the DM annihilation rate~\citep[e.g.][]{Gondolo2000,Bertone2002,Fields2014,Lacroix2017,Xia2021,Liu2022,Balaji2023}.

The formation of SMBHs is a complex process, and it may not always be adiabatic or initially located at the centre of the galaxy, leading to a flattening of the DM spike.
Several studies have demonstrated that, by relaxing the ideal assumptions, the slope of the spike may differ.
The slope of the spike can be weakened to 3/2 due to sufficient scatterings of DM particles with dense stellar populations~\citep{Gnedin:2003rj,Merritt2004,Shapiro2022}. 
The formation of SMBHs through an instantaneous gas collapse can result in a spike slope of 4/3~\citep{Ullio2001}. 
Major mergers of DM halos containing SMBHs can heat DM particles, producing a cusp with a slope of $0.5$ within about 10 pc~\citep{Merritt2002a}.
Additionally, the spike density can be weakened if the seed BH is massive enough and located off-centre~\citep{Ullio2001}. 
However, other processes such as chaotic orbits in triaxial halos~\citep{Merritt2004b} or gravo-thermal collapse for self-interacting DM~\citep{Ostriker2000} may enhance the spike. 
Still, the innermost history of the Galaxy is uncertain~\citep{Chen2022}, 
making it difficult to determine the extent of influence on the DM spike. 
Therefore, in order to better constrain the DM parameters, it is necessary to probe the spike in observations. 
There are two options: the stellar orbits of S-stars~\citep{Weinberg2005,Lacroix:2018zmg} and the gravitational wave~\citep{Eda2013,Li2022,Shen2023GW,Ghoshal2023}, while our work only focus on the former one.

Over the past three decades, considerable efforts have been made to accurately measure the stellar kinematics in the innermost region of the Galaxy~\citep{Eckart1996,Ghez1998,Schodel2002,Ghez2003,GRAVITY:2018ofz,Do:2019txf}.
Thanks to the high resolution of the Keck observatory and the Very Large Telescope (VLT), the orbits of more than 40 S-stars are currently available~\citep{Gillessen2017,Peissker2020b,Peissker2020a,Peissker2022,GRAVITY2021b}.
These data have significantly improved our understanding of the SMBH properties~\citep{Genzel2010,GRAVITY2022a,Zhang2015}, 
the environment around the black hole~\citep{GRAVITY2019,Bar2019,Becerra-Vergara2021,Benisty2022,Yuan2022b,GRAVITY2023}, 
and even the gravity theory~\citep{Hees:2017aal,GRAVITY2020a,Yan:2022fkr,DellaMonica2023}.
Recently, due to the updated orbit measurements of S-stars~\citep{Gillessen2017,Do:2019txf,GRAVITY2022a} and the first image of Sgr A* presented by the Event Horizon Telescope (EHT)~\citep{EHT2022}, interests in the DM spike are aroused.
In a study by \citet{Lacroix:2018zmg}, the VLT and Keck observations of S2 based on the period before 2016 were analyzed, 
and the size of the spike was constrained. 
Their results only exclude the GS spike model for $\gamma \gtrsim 1.4$ with 95\% probability 
because the star had not yet reached its pericentre at that time.
In a subsequent study by~\citet{GRAVITY2020a}, the radius of the spike for $\gamma = 1$ was further constrained by using the GRAVITY/VLT data. 
In another recent study by \citet{Nampalliwar:2021tyz}, the Keck measurement of S2 was used to determine the inner radius and density of the spike, with a particular focus on the impact of the spike on the EHT shadow image.

In this work, we revisit and extend the analyses of DM spike with the public data of the S-stars from Keck and VLT.
Our work not only updates the constraints on the radial extension parameter $R_{\rm sp}$ for different values of $\gamma$, 
but also establishes, for the first time, limits on the spike slope $\gamma_{\rm sp}$ using the S2 orbit data.
We further investigate the impact of combining data from multiple S-stars and   
calculate the Einasto profile as the initial density distribution.\footnote{ 
Hereafter, when referring to the \textit{initial density profile}, it means  
the halo profile before undergoing accretion by the black hole.}
We also discuss the requirement of the annihilation cross section for the NFW spike with a weak cusp to escape the constraints of the S2 orbit.

This paper is organized as follows.
In Sec.~\ref{sec::profile}, we introduce the density profiles of DM spike. 
In Sec.~\ref{sec::method}, we present the orbit data of the S-stars, 
the post-Newtonian dynamical model, and 
the statistical method employed in our analysis.
In Sec.~\ref{sec::results}, we show our constraints of the NFW spike and Einasto spike using the S2 orbital data from the Keck and VLT.
In Sec.~\ref{sec::discussion}, we further discuss the constraints when more S-stars are combined, when a full-orbit GRAVITY measurements are adopted, and when DM particles can annihilate.
Finally, we summarize our work in Sec.~\ref{sec::summary}.

\section{Dark Matter Profile}\label{sec::profile}
It is a popular assumption that a spike profile can be formed after the adiabatic growth of SMBH~\citep{Gondolo:1999ef}.  
Under such an adiabatic condition, 
one can obtain the analytical form of the NFW spike in the GS model assuming the conservation of angular momentum and radial action~\citep{Gondolo:1999ef,Young1980}. 
On the other hand, the analytical form of the Einasto spike in the GS model is hard to find, but  
one can obtain its circular-orbit approximation~\citep{Ullio2001,Blumenthal1986}.
In this section, we introduce the spike profiles:   
the NFW spike presented in Sec.~\ref{sec:nfw_spike} 
and the Einasto spike presented in Sec.~\ref{sec::profile:einasto}.

Note that we are also phenomenologically interesting the general case of removing the assumptions in the GS model. 
Except presenting the GS spike profiles for both halo models, 
we will also compare generic spike with the GS spike by releasing the conditions
and hence we can have one free parameter in the generic spike scenario.

\begin{figure}
\begin{center}
\includegraphics[width=\columnwidth]{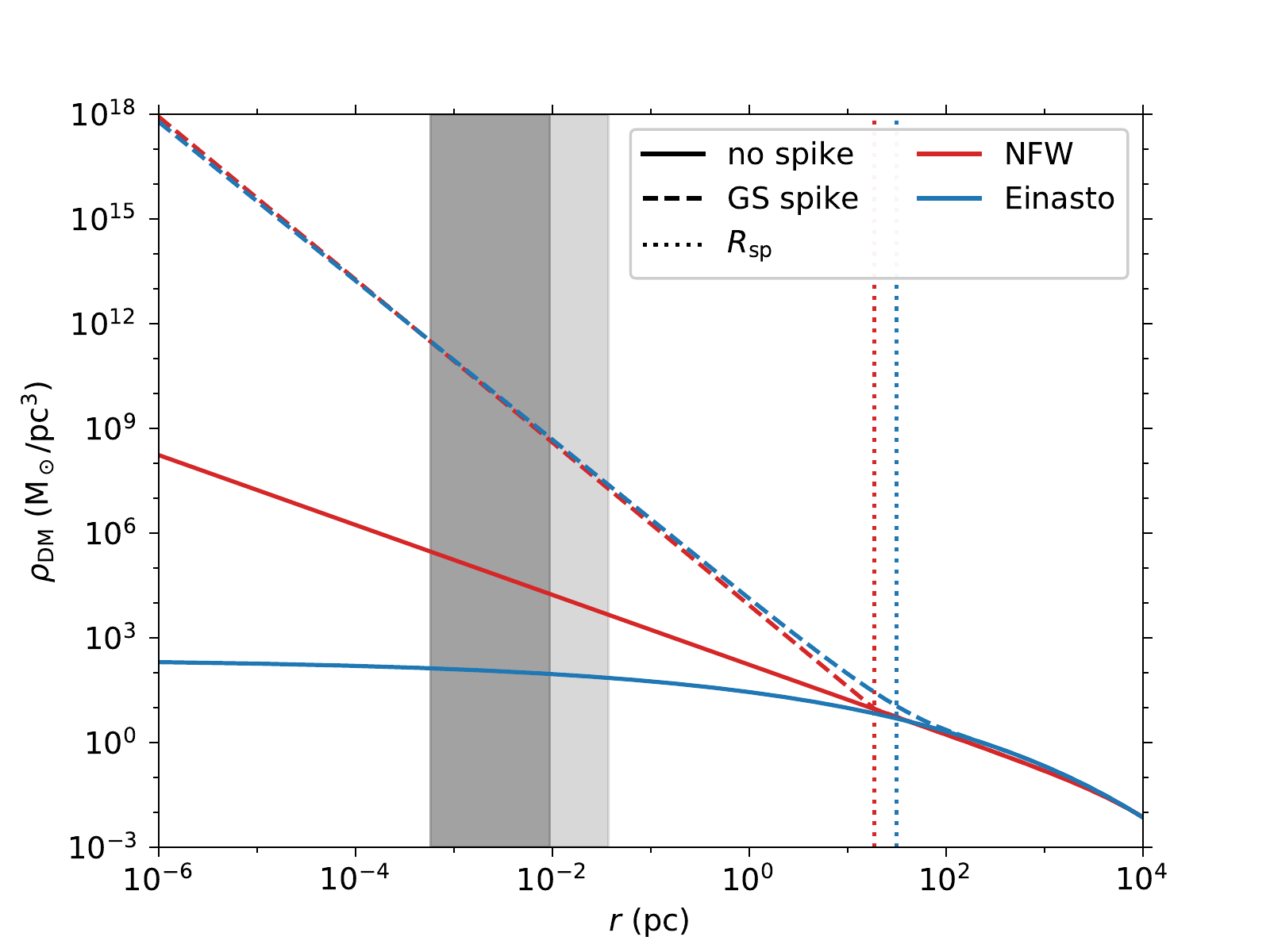}
\end{center}
\caption{\label{fig::profile}
The possible DM density profile in the Milky Way. 
The red and blue lines are for the initial NFW and Einasto profile, respectively.
The solid and dashed lines represent the profiles with and without the GS spikes.
The radial extension radii are also drawn in dots.
The ranges covered by S2-only and the four stars (S2, S1, S9 and S13) are indicated, respectively, by the dark shaded region and the entire shaded region. 
    }
\end{figure}

\subsection{The NFW spike}
\label{sec:nfw_spike}
The most popular DM density model is the generalized Navarro-Frenk-White (gNFW) profile.
The density at the Galactocentric radius $r$ is
\begin{equation}\label{eq::gnfw_halo}
    \rho_{\rm gnfw,halo}(r)=\frac{\rho_0}{(r/r_{\rm s})^\gamma (1+r/r_{\rm s})^{3-\gamma}}.
\end{equation}
Here, $r_{\rm s}$ is the scale radius with the condition $\dd \ln \rho/\dd \ln r=-2$.
The steepness of the density profile within $r_{\rm s}$ is defined by $\gamma$, which equals $1$ for the original NFW profile~\citep{Navarro1997}.
The density normalization $\rho_0$ can be determined from the observations~\citep{McMillan2017,Cautun2020,Benito2021,Wang2022}.
In our work, we adopt the gNFW profiles for different $\gamma$ from~\citet{McMillan2017} 
and mainly focus on the case of $0.5\leq \gamma \leq 1.5$.
The upper value of $\gamma$ covers the results given in \cite{McMillan2017} and \cite{Wang2022}, and the lower value is adopted simply due to that the GS model with $\gamma<0.5$ is hardly constrained by the current data of the S-stars and is beyond our interests.

For the general gNFW spike profile (no adiabatic assumption), 
we adopt the piece-wise function~\citep{Lacroix:2018zmg}
\begin{equation}\label{eq::nfw_combine}
    \rho_{\rm gnfw,sp}(r) =
    \begin{cases}
        0 & \qquad r \leq 2R_{\rm sch},\\
        \rho_{\rm sp0} \left( \frac{r}{R_{\rm sp}}\right)^{-\gamma_{\rm sp}} & \qquad 2R_{\rm sch} < r \leq R_{\rm sp},
    \end{cases}
\end{equation}
where $\rho_{\rm sp0}=\rho_{\rm gnfw,halo}(R_{\rm sp})$ and $R_{\rm sch}$ is the Schwarzschild radius of the GC SMBH which is $2GM_{\rm BH}/c^2$ with $M_{\rm BH}\approx 4\times 10^6~M_\odot$.
$R_{\rm sp}$ and $\gamma_{\rm sp}$ are the spike radius and slope, respectively, which are two independent parameters to determine the spike density.

If considering the spike after the adiabatic growth of SMBH (GS spike), its
spike slope is
\begin{equation}\label{eq::gs_spike_slope}
\gamma_{\rm sp}^{\rm GS}=(9-2\gamma)/(4-\gamma),
\end{equation}
and the spike radius is
\begin{equation}\label{eq::gs_spike_radius}
R_{\rm sp}^{\rm GS}=a_\gamma r_{\rm s} (M_{\rm BH}/(\rho_0 r_{\rm s}^3))^{1/(3-\gamma)},
\end{equation}
where $a_\gamma$ is the scale factor interpolated from the values in~\citet{Gondolo:1999ef}. 
We can see that $R_{\rm sp}^{\rm GS}$ and $\gamma_{\rm sp}^{\rm GS}$ both are the function of $\gamma$.
As a reference, the predicted GS spike radii and slopes are given in Tab.~\ref{tab::constraints}.
Upon comparison of the NFW halo profile with and without the spike profile, 
depicted by the red solid and dashed lines correspondingly in Fig.~\ref{fig::profile},  
a significant increase in the DM density of approximately six orders of magnitude can be observed at the pericentre of S2.

\subsection{The Einasto spike}\label{sec::profile:einasto}
The Einasto profile is also a commonly used DM distribution model.
By accounting for the power-law evolution of the logarithmic density slope with respect to the radius, 
this model can better fit the numeric simulations~\citep{Einasto1965,Navarro2004,Wang2020}.
It can be written as
\begin{equation}\label{eq::einasto_halo}
    \rho_{\rm ein,halo}(r)=\rho_{0} \exp \left \{-\frac{2}{\alpha}\left[\left(\frac{r}{r_{\rm s}}\right)^{\alpha}-1\right]\right \},
\end{equation}
where $\rho_0$ and $r_{\rm s}$ are the normalization and scale radius respectively, and $\alpha$ is the inverse of the Einasto index which characterizes the mass concentration.
In this work, we only choose two parameter benchmarks for the Einasto profile as a representation of different Einasto index and local DM density.
The first benchmark labelled as ``\einn'' is originated from the simulation~\citep{Navarro2004}: 
\{$\rho_{\rm ein,halo}({\rm 8.2~\rm kpc})=0.01~M_\odot\rm \,pc^{-3}$, $r_{\rm s}=20~\rm kpc$, $\alpha=0.17$\}. 
However, the second one labelled as ``\einw'' is derived by the rotation curve and globular cluster kinematics 
from {\it Gaia}~\citep{Wang2022}: \{$\rho_{\rm ein,halo}({\rm 8.2~\rm kpc})=0.008~M_\odot\rm \,pc^{-3}$, $r_{\rm s}=12~\rm kpc$, $\alpha=0.32$\}.

\begin{figure}
\includegraphics[width=\columnwidth]{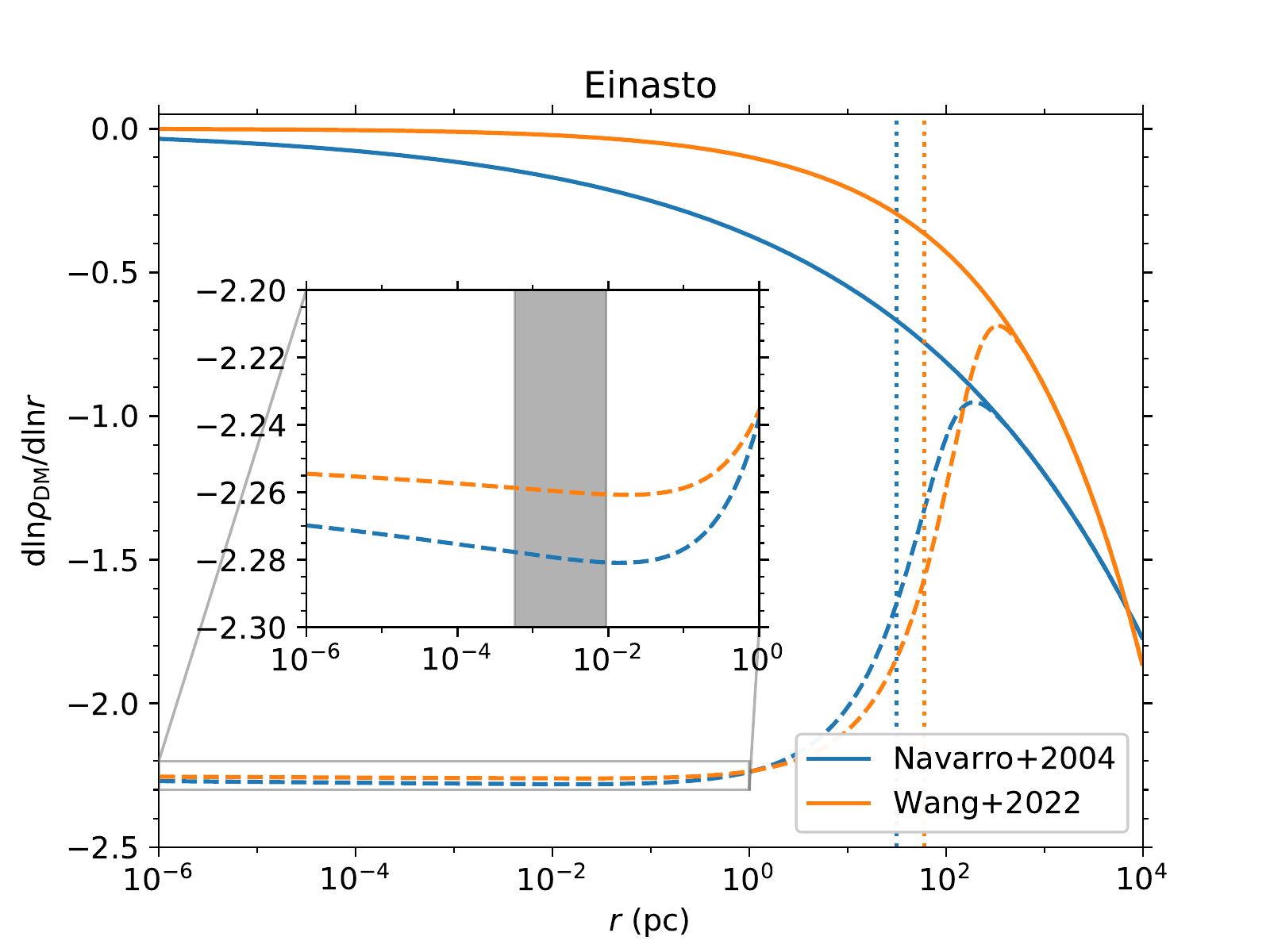}
\caption{\label{fig::einasto_index}
The logarithmic slope of the Einasto profile with and without the GS spike in solid and dashed lines respectively.
The blue and orange lines correspond to the Einasto profiles with the parameter sets labeled as \einn~\citep{Navarro2004} and \einw~\citep{Wang2022}, respectively. The inset shows the slopes of the GS spike profiles for the two models within $1~\rm pc$. The shaded region covers the orbit of S2.
}
\end{figure}

We adopt the circular-orbit approximation of the GS spike model~\citep{Blumenthal1986,Ullio2001}, which only differs from the semi-analytical one by no more than a factor of two~\citep{Ullio2001}.
The spike profile results from the conservation of the angular momentum, the radial action and the phase space distribution during the adiabatic growth of the BH~\citep{Young1980}.
In the circular-orbit approximation, the radial action is zero, the angular momentum conservation is given by
\begin{equation}\label{eq::einasto_am}
    r_i M_{{\rm tot},i}(r_i) = r_f M_{{\rm tot},f}(r_f),
\end{equation}
and the conservation of distribution implies
\begin{equation}\label{eq::einasto_dis}
    M_{{\rm dm},i}(r_i) = M_{{\rm dm},f}(r_f),
\end{equation}
where $M_{\rm tot}(r)$ and $M_{\rm dm}(r)$ are the enclosed total mass and DM mass inside the radius $r$ respectively, and the subscript $i$ and $f$ denote the initial and final state.
The the mass distribution of DM before the growth of SMBH are
\begin{equation}
M_{{\rm dm},i}(r) = \int_0^r 4\pi r^2 \dd r \rho_{\rm ein,halo}(r).
\end{equation}
Given the DM mass distribution after the growth $M_{{\rm dm},f}$, the spike density profile can be calculated with
\begin{equation}
    \rho_{\rm ein,sp}(r) =
    \begin{cases}
        0 & \qquad r \leq 2R_{\rm sch},\\
        \frac{1}{4\pi r^2} \frac{\dd M_{{\rm dm},f}}{\dd r} & \qquad 2R_{\rm sch} < r \leq R_{\rm sp},
    \end{cases}
\end{equation}
where the inner radius $2R_{\rm sch}$ is for the Schwarzschild BH~\citep{Sadeghian2013,Ferrer2017}.

For the GS spike, the SMBH accretes the interstellar medium from scratch in the centre, therefore $M_{{\rm tot},i}^{\rm GS}(r)=M_{{\rm dm},i}(r)$ and $M_{{\rm tot},f}^{\rm GS}(r)=M_{{\rm dm},f}^{\rm GS}(r)+M_{\rm BH}$.
The spike radii, defined with $M_{{\rm dm},f}^{\rm GS}(R_{\rm sp}^{\rm GS})=M_{\rm BH}$, are 31.2~pc and 60.4~pc for \einn and \einw profiles, respectively.

We depict the Einasto profile and the GS spike profile by the blue solid and dashed lines in Fig.~\ref{fig::profile}.
Although the Einasto profile is smaller than NFW profile by two orders of magnitudes in the S2 orbit region, the spike profile after accretion is quite similar.
The slopes of the GS spikes in the \einn and \einw benchmarks are presented by the blue and orange dashed lines in Fig.~\ref{fig::einasto_index}.
The spike slopes barely change within the spike radius and are equal to $2.26$ and $2.28$ for the \einn and \einw, respectively.
This can be understood that the phase space distributions of the Einasto profiles are singular around the BH~\citep{Cardone2005,Baes2022} so that the Einasto spikes are as steep as the gNFW spike with $\gamma=0$.

To release the GS assumptions, we choose the spike radius $R_{\rm sp}$ as the free parameter.
To obtain the enclosed mass profile given $R_{\rm sp}$, we use
\begin{equation}
    M_{{\rm tot},f}(r) = M_{{\rm dm},f}(r) + M_{\rm acc}(R_{\rm sp}),
\end{equation}
and
\begin{equation}
    M_{{\rm tot},i}(r) = M_{{\rm dm},i}(r),
\end{equation}
where $M_{\rm acc}(R_{\rm sp})$ is the mass of accreted interstellar medium, which is set to be $M_{\rm acc}(R_{\rm sp})=M_{{\rm dm},i}(2 R_{\rm sp})$.
With Eq.(\ref{eq::einasto_am}) and Eq.(\ref{eq::einasto_dis}), the enclosed mass profile $M_{{\rm dm},f}(r; R_{\rm sp})$ and thereby the density profile $\rho_{\rm ein,sp}(r;R_{\rm sp})$ can be derived.
If the spike radius $R_{\rm sp}^{\rm GS}$ is used, $M_{{\rm dm},i}(2R_{\rm sp}^{\rm GS})=M_{{\rm dm},f}(R_{\rm sp}^{\rm GS})=M_{\rm BH}$, therefore the enclosed mass profile $M_{{\rm dm},f}(r; R_{\rm sp}^{\rm GS})$ can reduce to the GS spike mass distribution.

\section{Analysis}\label{sec::method}
\subsection{Dynamical model}
To derive the stellar orbits, we solve the first-order post Newtonian approximation considering that the higher order effects are below the precision of current observation sensitivities~\citep{Do:2019txf}.
The equation of motion with a spherically symmetric mass distribution can be written as~\citep{Rubilar2001}
\begin{equation}\label{eq::motion_eq}
\frac{\dd^2 {\mathbfit r}}{\dd t^2} = 
    -\frac{G M_{\rm tot}(r)}{r^3}{\mathbfit r} 
    -\frac{G M_{\rm tot}(r)}{c^2 r^3}\left[ (4\phi(r)+v^2) {\mathbfit r} - 4{\mathbfit v} ({\mathbfit v}\cdot {\mathbfit r}) \right],
\end{equation}
where $M_{\rm tot}(r)=M_{\rm BH} + M_{\rm dm}(r)$ is the enclosed total mass which is related to the parameters of the spike.
$\phi(r)=-\int^r_\infty \dd r\,G M_{\rm tot}(r)/r^2$ is the gravitational potential at a given radius.
We define a coordinate system by setting the origin at the SMBH, letting the $X$ and $Y$ axes point to the west and north, and making the $Z$ axis point from the GC to the solar system~\citep{Do:2019txf,Yuan2022b}.
In this case, ${\mathbfit r}(t)\equiv [X(t), Y(t), Z(t))]$ and ${\mathbfit v}(t) \equiv {\dot {\mathbfit r}}(t) = [V_X(t), V_Y(t), V_Z(t)]$.

An initial condition is also required to solve the equation of motion.
We define the initial state for each star at the epoch $t_0=2000.0$ with the following six parameters:
the inclination $I$, the longitude of ascending node $\Omega$, the positions $(x_0,y_0)$ and velocities $(v_{x0}, v_{y0})$ in the orbital plane.
The initial phase-space coordinates can be transformed from the six parameters through
\begin{equation}\label{eq::initial_state}
    \begin{aligned}
    X(t_0)   &=  x_0 \cos\Omega - y_0 \cos I \sin\Omega , \\ 
    Y(t_0)   &=  x_0 \sin\Omega + y_0 \cos I \cos\Omega, \\ 
    Z(t_0)   &= -y_0 \sin I, \\
    V_X(t_0) &=  v_{x0} \cos\Omega - v_{y0} \cos I \sin\Omega, \\
    V_Y(t_0) &=  v_{x0} \sin\Omega + v_{y0} \cos I \cos\Omega, \\
    V_Z(t_0) &= -v_{y0} \sin I.
    \end{aligned}
\end{equation}
The {\tt DOP853} algorithm~\citep{DOP853} in {\tt SciPy} is utilized to solve the state of the star at epoch $t$.

The orbital planes of the S-stars are not perpendicular to the line of sight, so there is a time delay caused by the propagation of light through the orbit plane in $Z$ direction, i.e. the R{\o}mer delay.
This effect is detectable in the Keck observation of S2, which leads to a time delay of $-0.5$ days at pericentre and $7.5$ days at apocentre~\citep{Do:2019txf}.
The time delay is $t_{\rm obs} = t_{\rm em} - Z(t_{\rm em})/c$,
where $t_{\rm obs}$ and $t_{\rm em}$ represent the epochs of observation and emission respectively.
The epoch of emission can be solved with iterations~\citep{Hees2014}, however one iteration is adequate at the present~\citep{Do:2019txf}
\begin{equation}\label{eq::romer_delay}
    t_{\rm em}=t_{\rm obs}+Z(t_{\rm obs})/c.
\end{equation}
The second term has an opposite sign compared to~\citet{Do:2019txf}, since the $Z$ axis in their work is pointing from the Sun to the GC.
The Shapiro time delay is ignored in our work since the correction is merely $\lesssim 5~\rm min$.

Once we have the star position at any epoch, the right ascension (R.A.), declination (Dec.) and radial velocity can be calculated with
\begin{equation}\label{eq::celestial_rv}
    \begin{aligned}
        \alpha_*(t_{\rm obs}) &= {Y(t_{\rm em})}/{R_0} + \alpha_{\rm BH} + v_{\alpha,{\rm BH}}\cdot(t_{\rm em}-t_{\rm ref}), \\
        \delta_*(t_{\rm obs}) &= {X(t_{\rm em})}/{R_0} + \delta_{\rm BH} + v_{\delta,{\rm BH}}\cdot(t_{\rm em}-t_{\rm ref}), \\
        v_{r*}(t_{\rm obs})   &= V_Z(t_{\rm em}) + [ {V^2(t_{\rm em})}/{2} - \phi(r(t_{\rm em})) ]/c +v_{r0},
    \end{aligned}
\end{equation}
where $R_0$ is the distance between the Sun and the central SMBH and $t_{\rm ref}$ is the reference epoch which is 2000.0 for Keck~\citep{Do:2019txf} and 2009.0 for VLT~\citep{Gillessen2017}.
$(\alpha_{\rm BH}, \delta_{\rm BH})$ and $(v_{\alpha,{\rm BH}}, v_{\delta,{\rm BH}})$ are the offset and linear drift of the central mass in celestial coordinate.
$V=|{\mathbfit v}|$ is the norm of the velocity, and $v_{r0}$ is a constant velocity offset in the radial velocity measurements.

\subsection{Data and statistical method}
We use the latest publicly available astrometric and spectroscopic measurements of the S-stars from the VLT and Keck telescopes compiled in~\citet{Do:2019txf} and~\citet{Gillessen2017}.
\citet{Do:2019txf} presents the measurements of S2 from Keck/NIRSPEC, Keck/NIRC, Keck/NIRC2, Keck/OSIRIS, Gemini/NIFS, Subaru/IRCS and VLT/SINFONI between 1995 and 2019.
\citet{Gillessen2017} provides the latest publicly available measurements of 17 S-stars observed by NTT/SHARP, VLT/SINFONI, VLT/SPIFFI and VLT/NACO from 1992 to 2017.
Since the radial velocities compiled in~\citet{Do:2019txf} contain the VLT/SINFONI measurements in~\citet{Gillessen2017}, we remove the overlaps in the combined analysis.
Different offsets and drifts of the central object are adopted as free parameters for two observatories to explain their discrepancies in the results~\citep{Gillessen:2009ht,GRAVITY2021}, which enables us to combine the data sets.
For the S2, we have 190 astrometric points and 115 radial velocities with a full coverage of the orbit.
Although there are updated measurements of S2, S29, S38 and S55 from the VLT/GRAVITY interferometric and VLT/SINFONI spectroscopic observations~\citep{GRAVITY:2018ofz,GRAVITY2020a,GRAVITY2022a}, these data are not publicly available yet.

The total likelihood function is constructed as follows:
\begin{equation}\label{eq::likelihood_total}
\mathcal{L}_{\rm tot} = \mathcal{L}_{\rm astro,Keck} \times \mathcal{L}_{\rm astro,VLT} \times \mathcal{L}_{\rm rv},
\end{equation}
where $\mathcal{L}_{\rm astro,Keck}$ and $\mathcal{L}_{\rm astro,VLT}$ are the likelihood functions of the astrometric data from Keck and VLT respectively, $\mathcal{L}_{\rm rv}$ is the likelihood function for the radial velocity.
The total likelihood contains the following free parameters:
\begin{enumerate}
    \item 
    Two parameters concerning the SMBH mass $M_{\rm BH}$ and the distance to the GC $R_0$.
    \item 
    The offset ($\alpha_{\rm BH}$, $\delta_{\rm BH}$) and the drift ($v_{\alpha,{\rm BH}}$, $v_{\delta,{\rm BH}}$) of the central object in the celestial coordinate for each telescope.
    There are eight free parameters.
    \item 
    The velocity offset $v_{r0}$ and the additional systematic offset $v_{\rm offset}$ between the Keck/NICR2 and other instruments~\citep{Chu2018,Do:2019txf}.
    \item 
    Six parameters defining the initial state of each star: $x_0$, $y_0$, $v_{x0}$, $v_{y0}$, $I$, and $\Omega$.
    \item 
    One parameter on the DM spike: either the spike radius $R_{\rm sp}$ or the density slope of the spike $\gamma_{\rm sp}$.
    \item 
    Two parameters accounting for the astrometric correlation $(\Lambda, p)$.
\end{enumerate}
When $n$ stars are involved, there are $15+6n$
free parameters in the likelihood function.

For the astrometric data from the Keck, the correlation caused by the faint source confusion~\citep{Plewa2018} is considered as recommended in~\citet{Do:2019txf}.
Therefore the likelihood function of the astrometric data from Keck reads:
\begin{equation}
    \begin{aligned}
        {\mathcal L}_{\rm astro,Keck} \propto & (\left|{\boldsymbol\Sigma}_\alpha\right| \left|{\boldsymbol\Sigma}_\delta\right|)^{-\frac{1}{2}}\\
        &\times \exp \left[-\frac{1}{2} \left(\Delta {\boldsymbol \alpha}^T {\boldsymbol\Sigma}_\alpha^{-1} \Delta {\boldsymbol \alpha}+\Delta {\boldsymbol \delta}^T {\boldsymbol\Sigma}_\delta^{-1} \Delta {\boldsymbol \delta} \right) \right],
    \end{aligned}
\end{equation}
where $\Delta {\boldsymbol \alpha} \equiv \{\alpha_i - \mu_{\alpha}(t_{i})\}$
and $\Delta {\boldsymbol \delta} \equiv \{\delta_i - \mu_{\delta}(t_{i})\}$
are the vector differences between the observed and predicted astrometric data from the Keck.
The covariance matrices are
$[{\boldsymbol\Sigma}_\alpha]_{ij} \equiv \sigma_{\alpha_{i}} \sigma_{\alpha_{j}} \boldsymbol{\rho}_{i j}$
and $[{\boldsymbol\Sigma}_\delta]_{ij} \equiv \sigma_{\delta_{i}} \sigma_{\delta_{j}} \boldsymbol{\rho}_{i j}$, 
where $\sigma_{\alpha}$ and $\sigma_{\delta}$ are the uncertainties of the astrometric data, and $\rho$ is the correlation matrix defined as $[\boldsymbol{\rho}]_{i j}=(1-p) \delta_{i j}+p \exp \left[-d_{i j}/\Lambda\right]$,
where $d_{ij}$ is the angular distance between the two Keck data points $i$ and $j$.
We also set free the correlation length scale $\Lambda$ and the mixing parameter $p$ in the model.
On the other hand, we simply use the $\chi^2$ form of the likelihood function for the VLT astrometric data~\citep{Gillessen2017}
\begin{equation}\label{eq::chi2_astro_VLT}
    -2 \ln {\mathcal L}_{\rm astro,VLT} = \sum_j \left[
        \left( \frac{\alpha_{j}-\mu_{\alpha}(t_j)}{\sigma_{\alpha j}} \right)^2
      + \left( \frac{\delta_{j}-\mu_{\delta}(t_j)}{\sigma_{\delta j}} \right)^2
      \right],
\end{equation}
considering different noise models only change the parameters within the statistical uncertainty for the data set~\citep{GRAVITY2021}.

The likelihood function for the radial velocities is
\begin{equation}\label{eq::chi2_rv_VLT}
    -2 \ln {\mathcal L}_{\rm rv} = \chi^2_{\rm rv} = \sum_k \left( \frac{v_{r,k}-\mu_{v_r}(t_k)}{\sigma_{v_r,k}} \right)^2,
\end{equation}
where $v_{r,k}$ and $\mu_{v_r}(t_k)$ are observed and predicted radial velocity, and $\sigma_{v_r,k}$ is the uncertainty.

The Bayesian inference method is chosen in this work.
Two different approaches are adopted to analyze the spike profile.
In the first approach, we fix the spike slope $\gamma_{\rm sp}$ to the value in the GS model and constrain the spike radius $R_{\rm sp}$.
We assume a flat prior for $R_{\rm sp}$ over a range greater than $2R_{\rm sch}$~\citep{Lacroix:2018zmg} in this case.
In the second approach, we fix the spike radius $R_{\rm sp}$ to the GS predicted value and constrain the spike slope $\gamma_{\rm sp}$.
A prior distribution for $\gamma_{\rm sp}$ is assumed based on the flat prior of the enclosed DM halo mass within the S2 apocentre $r_{\rm apo}$, namely
\begin{equation}\label{eq::prior_gammasp}
    \pi(\gamma_{\rm sp}) = \frac{\dd M_{\rm dm}}{\dd \gamma_{\rm sp}} \pi(M_{\rm dm})
    \propto \frac{1 - (3-\gamma_{\rm sp})\ln (x_{\rm apo}) }{(3-\gamma_{\rm sp})^2}\times x_{\rm apo}^{3-\gamma_{\rm sp}},
\end{equation}
where $x_{\rm apo} \equiv r_{\rm apo}/R_{\rm sp}$ and $\gamma_{\rm sp}<3$.
The priors of the remaining parameters are set to be flat in linear space.

\begin{figure*}
    \includegraphics[width=0.85\textwidth]{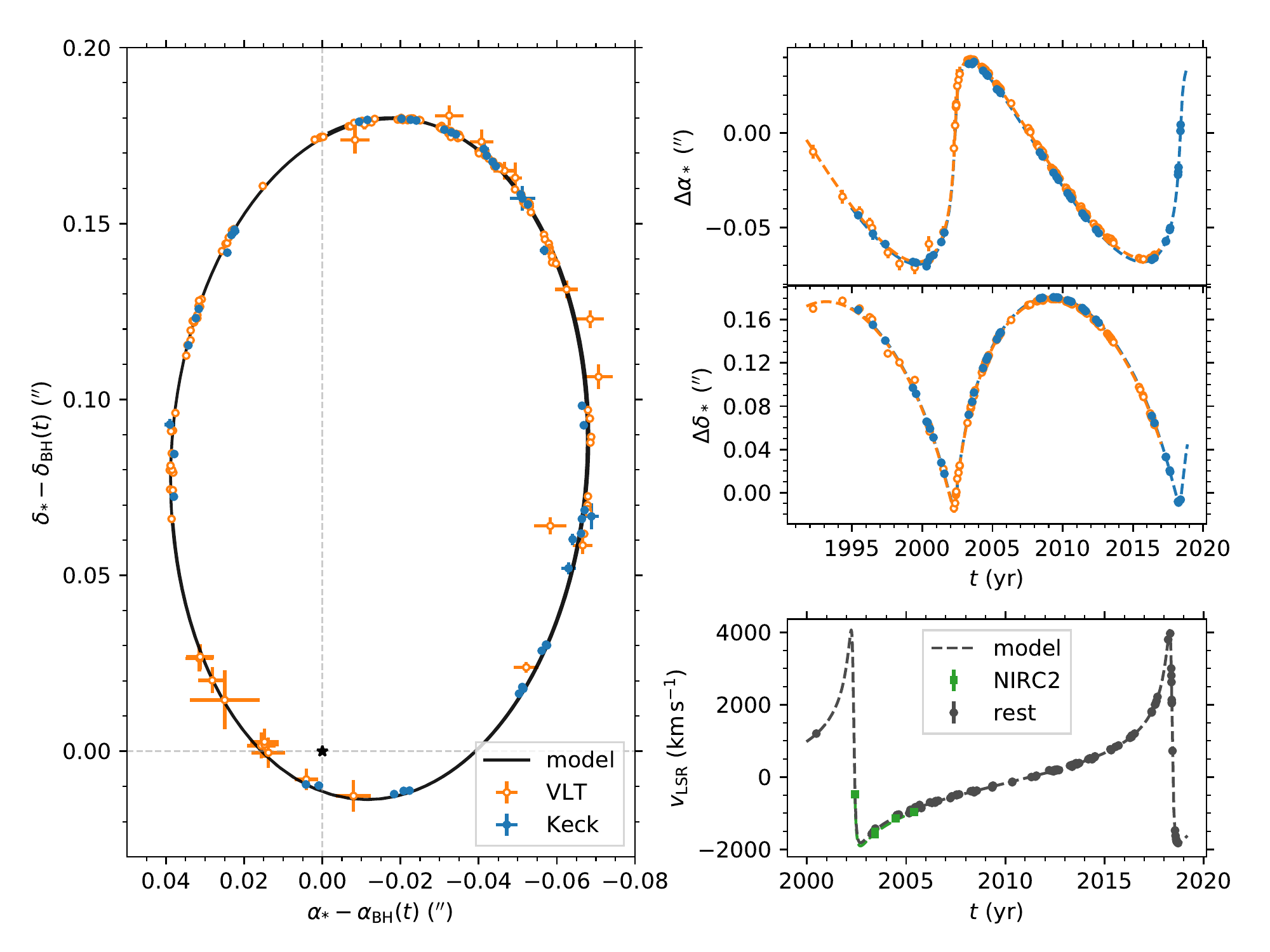}
    \caption{\label{fig::s2_bestfit}
        The data and the maximum-a-posteriori (MAP) model of S2 orbit around the SMBH with the DM spike of $\gamma=1$ included ($R_{\rm sp}$ is free).
        In the left panel, astrometric measurements corrected for the offsets of the reference frames observed by the VLT (Keck) are marked with orange hollow (blue solid) points.
        The black solid line shows the MAP orbit with the S2 moving clockwise.
        The Sgr A* is located at the origin marked with a black star.
        The right ascensions, declinations and radial velocities as a function of the epoch are shown in the upper, middle and lower right panels.
        The measurements are marked with points, while the optimal models are drawn with the dashed lines.
        Since there is a velocity offset for the Keck/NIRC2 data comparing to the rest data, we plot them separately in green square points.
    }
\end{figure*}

We adopt the Markov chain Monte Carlo (MCMC) sampler {\tt emcee}~\citep{emcee2013} to sample the posterior probability.
To ensure the samples well converged, we require the Gelman-Rubin diagnostic of $R-1 \lesssim 0.05$ and the samples longer than $\sim 30$ times the integrated autocorrelation per chain~\citep{Hogg2018,Cowles1996}.

\section{Results}\label{sec::results}
\subsection{Constraint on the generalized NFW spike with S2}\label{sec::gnfw}

The gNFW spike profile contains two parameters, the spike radius $R_{\rm sp}$ and the spike slope $\gamma_{\rm sp}$, which can be altered due to the massive seed BH or the dynamical heat process.
Since the S2 orbit only covers a small range of the spike (the dark grey band in Fig.~\ref{fig::profile}), it is difficult to determine the two parameters at the same time.
Hence, in this subsection, we treat the upper limits of $R_{\rm sp}$ and $\gamma_{\rm sp}$ separately, and keep the other parameter fixed to the value given in the GS model.

\subsubsection{The spike radius $R_{\rm sp}$}\label{sec::gnfw:rsp}
\begin{figure*}
    \includegraphics[width=\columnwidth]{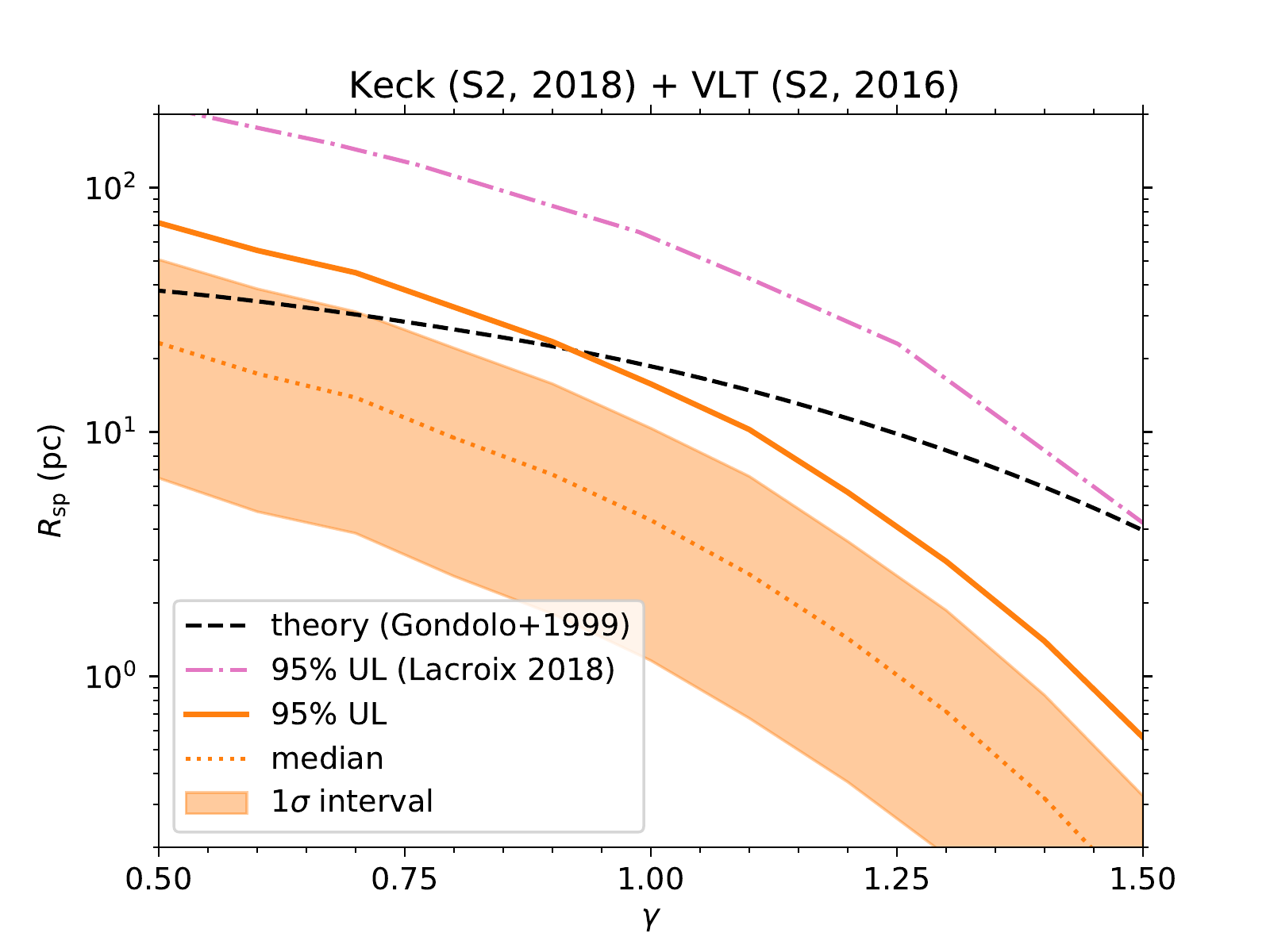}
    \includegraphics[width=\columnwidth]{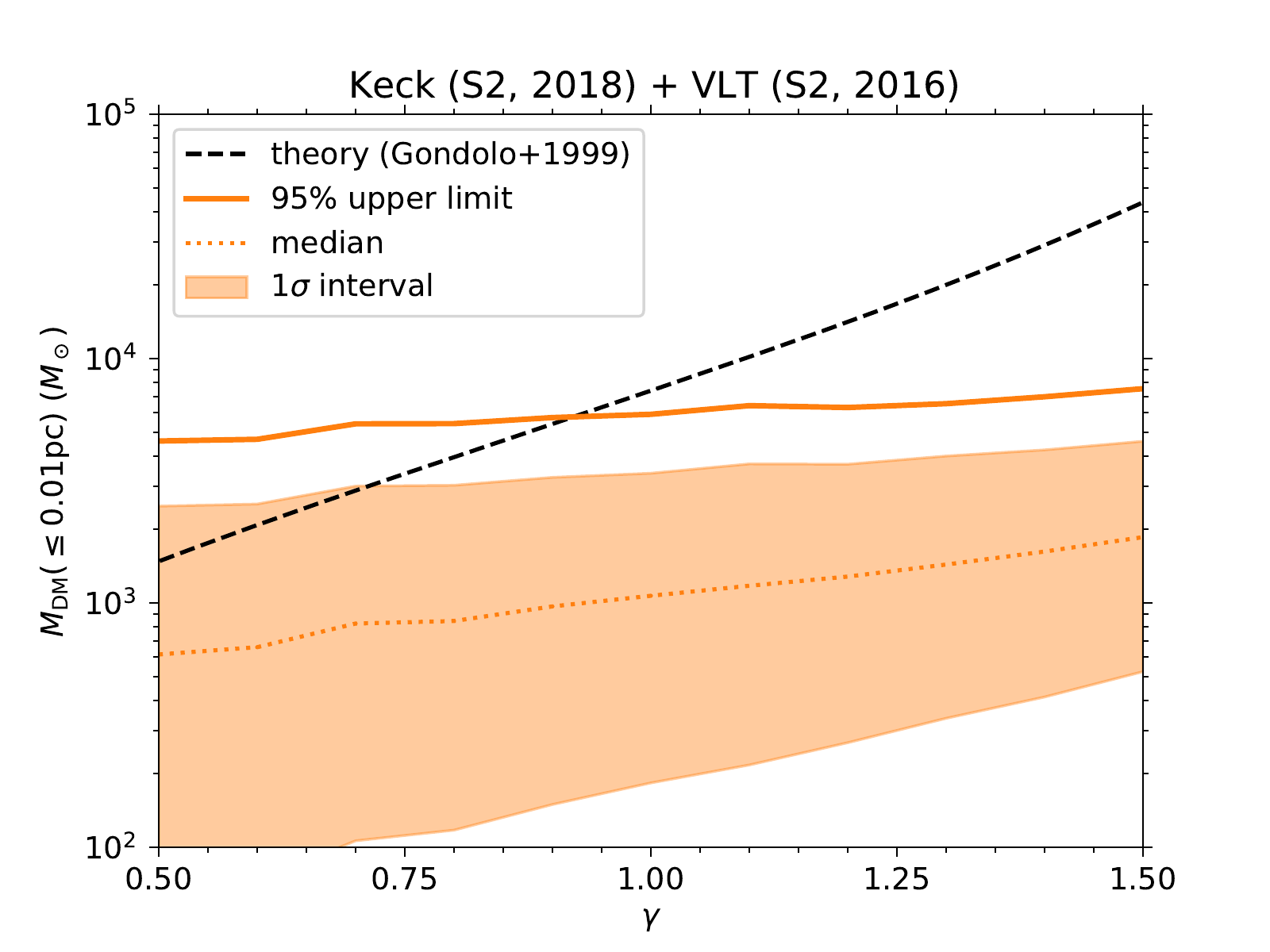}
    \caption{\label{fig::gnfw_Rsp_s2}
        The constraints of the gNFW spike profile with the photometry and velocity measurements of S2 from Keck and VLT.
        The spike slope $\gamma_{\rm sp}$ is fixed to the GS model value.
        The medians (orange dotted line), 68\% credible intervals (orange band) and 95\% upper limits (orange solid line) of the spike radius $R_{\rm sp}$ (left) and enclosed mass within $0.01~\rm pc$ (right) are illustrated.
        The upper limit from~\citet{Lacroix:2018zmg} is shown in the pink dot-dashed line.
        The theoretical GS spike model (black dashed line) is not favored by data when $\gamma \geq 0.92$ by 95\% probability.
    }
\end{figure*}

\begin{figure}
    \includegraphics[width=\columnwidth]{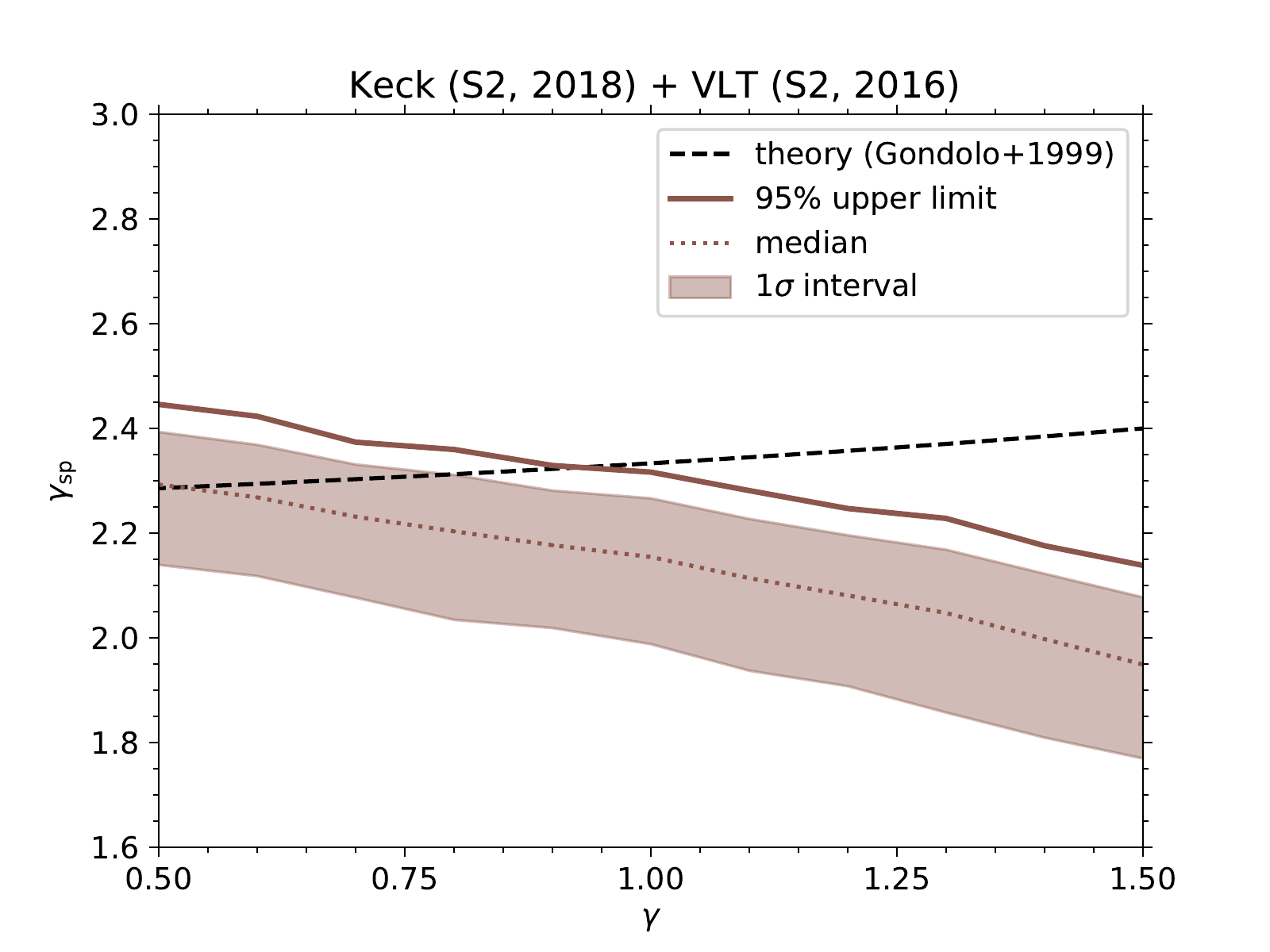}
    \caption{\label{fig::gnfw_gammasp}
        The constraints of the gNFW spike slope $\gamma_{\rm sp}$ with the photometry and velocity measurements of S2 from Keck and VLT.
        The spike radius $R_{\rm sp}$ is fixed to the GS model value.
        The corresponding medians (orange dotted line), 68\% credible intervals (orange band) and 95\% upper limits (orange solid line) are shown.
        The GS spike model (black dashed line) is not favored by data when $\gamma \geq 0.92$ by 95\% chance.
    }
\end{figure}

We analyze the combined orbital data of S2 with the gNFW halo model including a spike.
In this part, we constrain the spike radius $R_{\rm sp}$ and fix the spike slope to the GS predicted value.
In Fig.~\ref{fig::s2_bestfit}, we plot the measurements of S2 overlaid with the maximum-a-posteriori (MAP) spike model with $\gamma=1$.
The left panel is the celestial coordinates from the SMBH corrected for the offsets and drifts of the two telescopes.
The orange hollow and blue solid points are the astrometric measurements from the VLT and Keck telescopes respectively, while the black line represents the MAP orbit of S2 which moves clockwise.
The top, middle and bottom panels on the right are the R.A., Dec. offsets and radial velocity as a function of time.
The MAP model can well fit the measurements.

Following~\citet{Lacroix:2018zmg}, we take into account different gNFW density slopes $\gamma$ and analyze the constraint on the spike radius $R_{\rm sp}$ for each case.
In the left panel of Fig.~\ref{fig::gnfw_Rsp_s2}, we show the 68\% credible interval and 95\% upper limit of $R_{\rm sp}$ with the colour band and solid line.
The spike spatial extensions larger than 71.9~pc, 15.7~pc and 0.56~pc for the gNFW slope $\gamma$ of 0.5, 1.0 and 1.5 are excluded at 95\% level.
The $R_{\rm sp}$ constraint for a steeper initial slope $\gamma$ tends to be more stringent since it causes more DM mass within the range of S2 orbit.
We also draw the 95\% upper bounds from~\citet{Lacroix:2018zmg} with a pink dot-dashed line in the figure.
Our constraints are much stronger thanks to accurate Keck measurements of S2 at the closest approach.
The predicted spike radius in the GS model (Eq.(\ref{eq::gs_spike_radius})) is shown in the black dashed line and the ones with $\gamma \geq 0.92$ are excluded at 95\% level.
Even when we consider the systematic uncertainty of the local DM density $\rho_{\rm halo, \odot} = 0.008 - 0.013~M_{\odot}\rm \,pc^{-3}$~\citep{deSalas2021}, the predicted spike radius for the initial gNFW with $\gamma=1$ ranges from 15.8~pc to 20.5~pc and is still inconsistent with data.

We also convert the $R_{\rm sp}$ to the enclosed DM spike mass within the apocentre of S2.
As shown in the right panel of Fig.~\ref{fig::gnfw_Rsp_s2}, the constraint only weakly depends on $\gamma$,
confirming that the combined orbital data of S2 is not sensitive to the slope of the distribution at the present~\citep{Do:2019txf}.
The 95\% credible upper limit of enclosed mass is found to be $\sim 6000~M_\odot$, when the combined S2 data are adopted.
The extended mass constraint is weaker than that of the GRAVITY~\citep{GRAVITY2020a}, since the latter contains the private interferometric measurements which are more accurate than the adaptive optics.

The constrained spike models are consistent with other types of observations.
Such a spike increases the rotation velocity of gas by $\lesssim 2\%$ within the Galactocentric radius $\lesssim 5~\rm pc$, which is much smaller than the statistical uncertainties of the rotation curve ($15\%-30\%$)~\citep{Sofue2013}.
The spike contributes to an extended mass of $\lesssim 20~M_\odot$ within $2\times 10^{-6}~\rm pc$, which is not in conflict with the measurements of three hot spots around the SMBH considering the uncertainty of the inferred enclosed mass of $\sim 10^6~M_\odot$~\citep{GRAVITY2018b}.
The median precession angle of S2 is $10'-11'$ per orbit when such a spike exists (Fig.~\ref{fig::s2prec}), consistent with the GRAVITY measurements within $\sim 1\sigma$ uncertainty~\citep{GRAVITY2020a}.

\subsubsection{The spike slope $\gamma_{\rm sp}$}\label{sec::gnfw:gammasp}
We analyze the spike slope $\gamma_{\rm sp}$ using the measurements of S2.
The spike radius is fixed to the predicted value in the GS model.
The same as the previous part, we test the gNFW models with different halo density slope $\gamma$.
In Fig.~\ref{fig::gnfw_gammasp}, we present the 68\% credible interval and 95\% upper limit of $\gamma_{\rm sp}$ with the colour band and solid line.
The spike slope $\gamma_{\rm sp}$ steeper than 2.45, 2.32 and 2.14 for the $\gamma$ of 0.5, 1.0 and 1.5 are disfavoured by the S2 orbit at 95\% probability.
We draw the spike slopes predicted by the GS spike model (Eq.(\ref{eq::gs_spike_slope})) in the black dashed line.
The GS spike with $\gamma \geq 0.92$ is excluded at 95\% level.

The DM spike slope $\gamma_{\rm sp}$ could be flattened to 3/2 when the DM particles are efficiently heated by the dynamical processes~\citep{Merritt2002a,Gnedin:2003rj,Shapiro2022}.
The DM spike profile may also resemble the star spike profile in the GC whose slope $\gamma_{\rm sp}$ is $1.1-1.6$~\citep{Schodel2020,GRAVITY2020a}, considering the collisionless nature of DM particles.
However, it is difficult to constrain such possibilities with the current S2 measurements at the present.

\begin{figure}
    \includegraphics[width=\columnwidth]{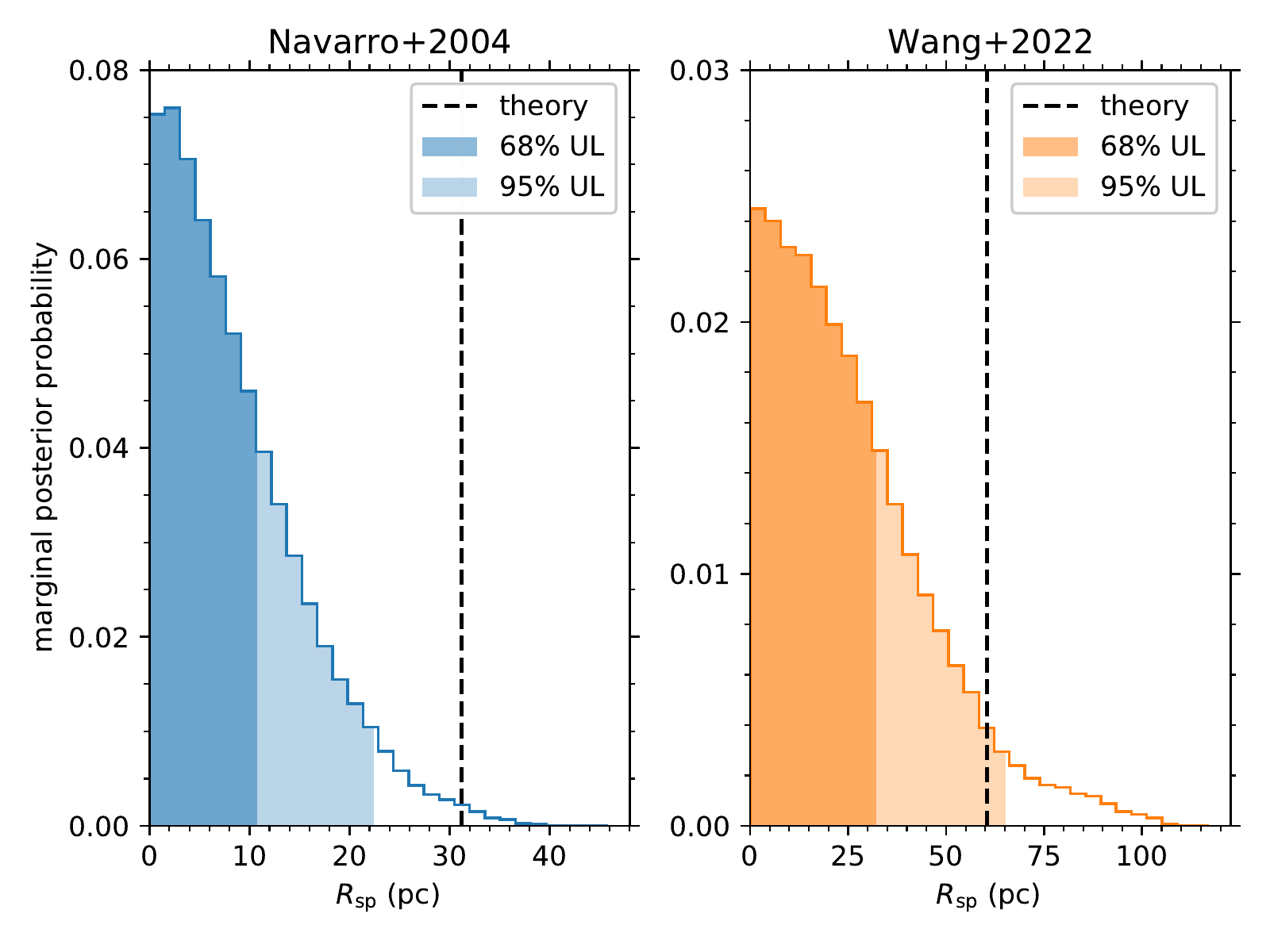}
    \caption{\label{fig::ein_Rsp_s2}
        The marginal posterior distribution of the spike radius $R_{\rm sp}$ of the Einasto spike.
        The left and right panels correspond to the Einasto spike profiles labeled as \einn and \einw.
        The dark and light shaded regions represent the 68\% and 95\% credible upper limits.
        The dashed line shows the spike radius which results from the adiabatic growth of the SMBH $M_{\rm BH}$ as given in Sec.~\ref{sec::profile}.
        The spike profile initiated from the profile \einn is disfavored by 95\% probability.
    }
\end{figure}

\begin{figure*}
    \includegraphics[width=0.81\textwidth]{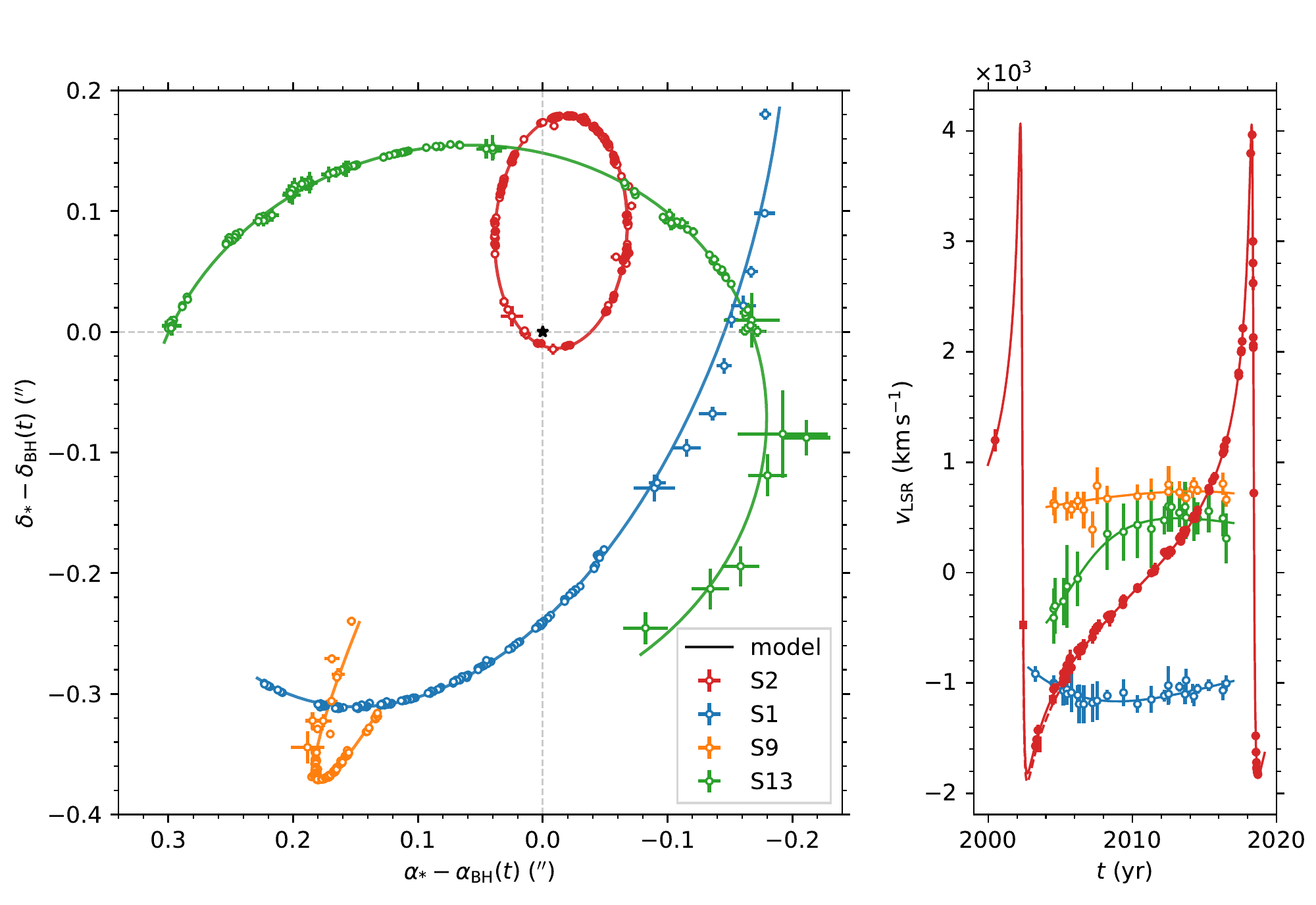}
    \caption{\label{fig::4stars_bestfit}
        The measurements and the MAP models of S2, S1, S9 and S13 around the SMBH with the DM spike of $\gamma=1$ included ($R_{\rm sp}$ is free).
        The left panel shows the astrometric points corrected for the offsets of reference frames, while the right panel shows the radial velocities for the epoch of observation.
        The hollow and solid points represents the data from the VLT and Keck respectively.
        The solid lines illustrate the MAP orbits and radial velocities.
        S2 moves clockwise, while S1, S9 and S13 move anti-clockwise.
    }
\end{figure*}

\begin{figure}
    \includegraphics[width=\columnwidth]{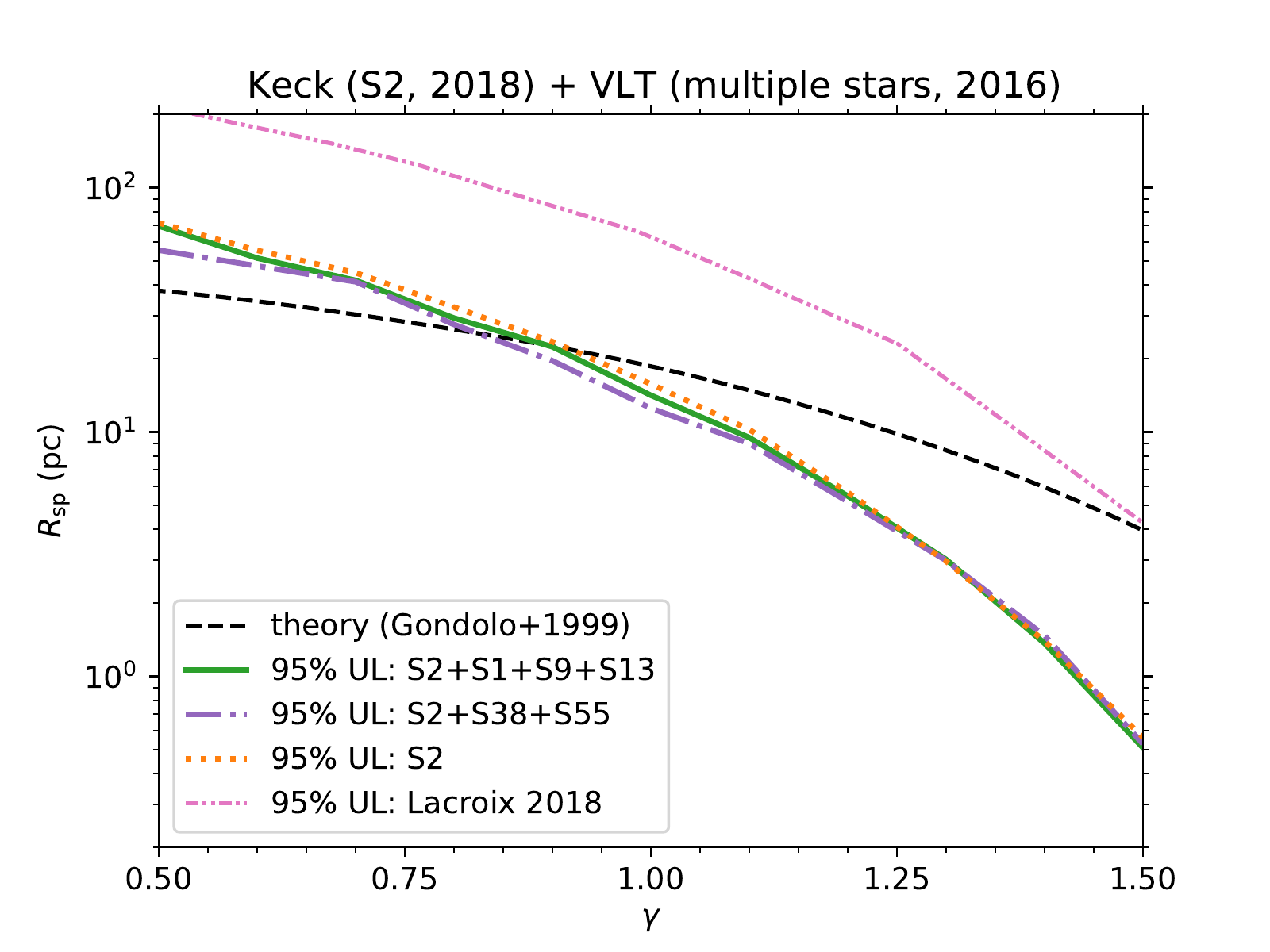}
    \caption{\label{fig::gnfw_Rsp_4stars}
        The constraints of the gNFW spike radius $R_{\rm sp}$ with the measurements of S2 from Keck and multiple S-stars from VLT.
        The spike slope $\gamma_{\rm sp}$ is fixed to the value in the GS model.
        The green solid, purple dot-dashed, and orange dotted lines correspond to the 95\% upper limits given the data set of (S2, S1, S9, S13), (S2, S38, S55), and S2, respectively.
        The 95\% upper limit from~\citet{Lacroix:2018zmg} is also shown in the pink dot-dot-dashed line.
        The GS spike model (black dashed line) is not favored by data when $\gamma \geq 0.83$ by 95\% chance.
    }
\end{figure}

\subsection{Constraint on the Einasto spike with S2}\label{sec::einasto}
The Einasto profile is among the most popular models for the DM density profile, however, the corresponding spike profile has not been discussed and constrained in the literature yet.
Here we set upper limits on the spike radius in the two parameter benchmarks of the Einasto spike profile given in Sec.~\ref{sec::profile:einasto}.

Fig.~\ref{fig::ein_Rsp_s2} presents the marginal posterior probability of spike radius $R_{\rm sp}$ for two Einasto parameter sets.
The dark and light shaded regions represent the 68\% and 95\% credible upper limits.
For the benchmark \einn and \einw, the 95\% upper limits of the spike radius are 21.5~pc and 61.4~pc respectively.
Comparing to the expected spike radii of 31.2~pc and 60.4~pc in the GS model shown in the black dashed lines, 
the steep \einn spike is disfavored at 95\% level and 
the flat \einw spike marginally survives from the S2 constraint.

\section{Discussion}\label{sec::discussion}
\begin{figure*}
    \includegraphics[width=\columnwidth]{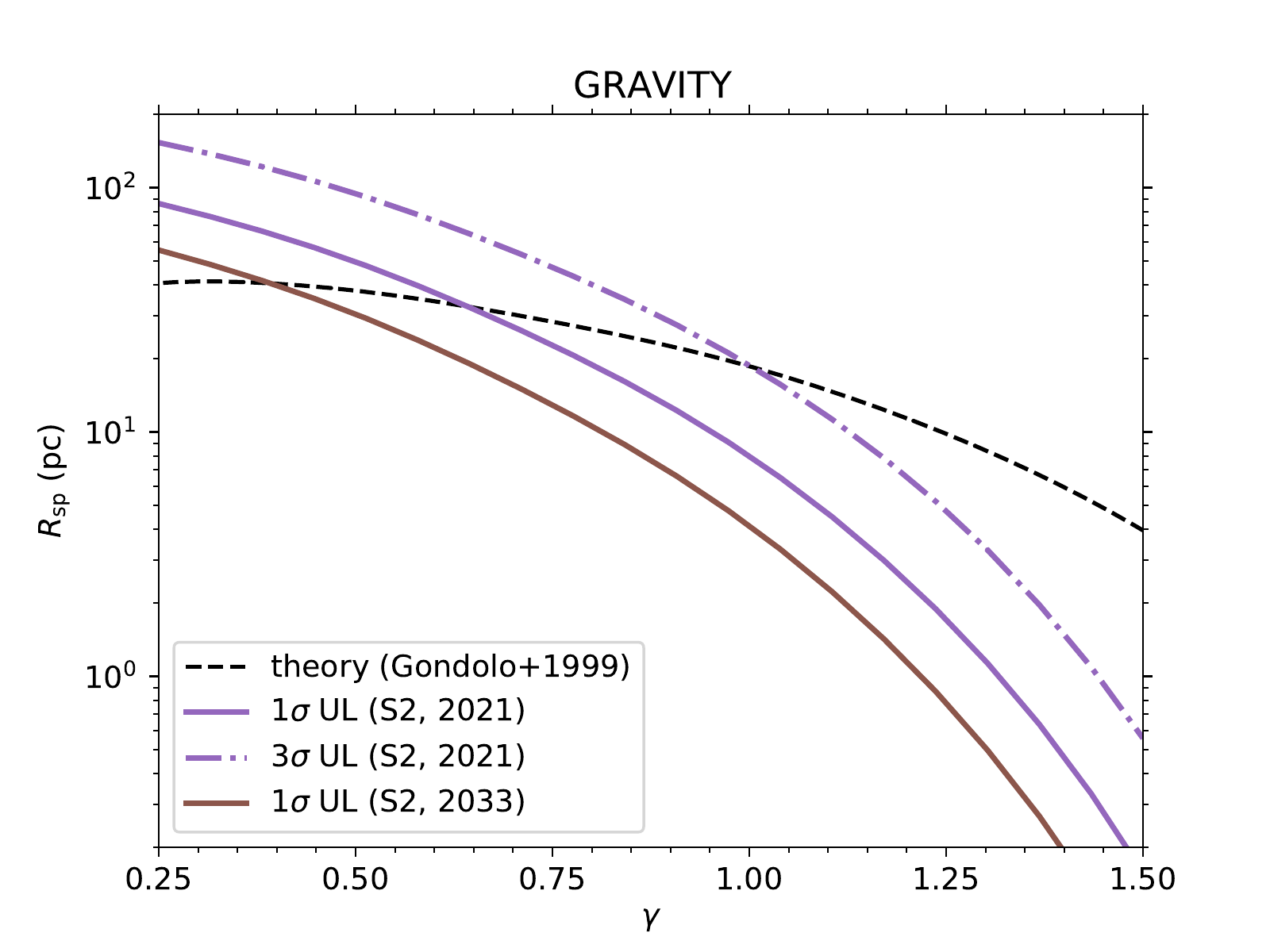}
    \includegraphics[width=\columnwidth]{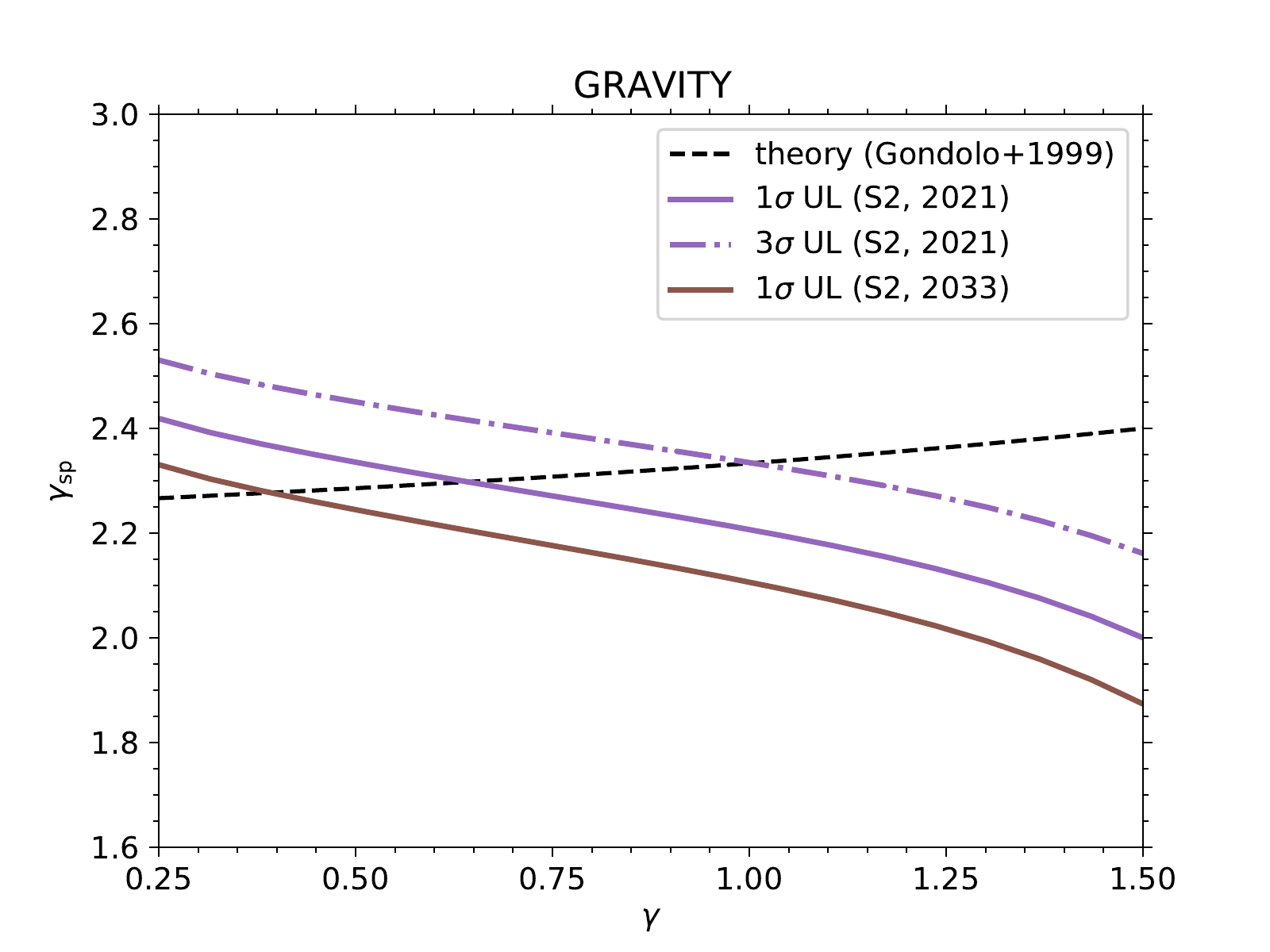}
    \caption{\label{fig::gnfw_gravity}
        The upper limits (ULs) on the spike radius $R_{\rm sp}$ (left) and slope $\gamma_{\rm sp}$ (right) of the gNFW spike profile converted from the GRAVITY upper limits of the extended mass within the apocentre of S2.
        The solid and dot-dashed purple lines come from the $1\sigma$ and $3\sigma$ constraints up to 2021 in~\citep{GRAVITY2022a}, while the brown solid line is from the expected $1\sigma$ constraint from a full orbit of the GRAVITY data~\citep{Heissel2022}.
        The expected spike radius in GS spike model is shown with the black dashed line.
    }
\end{figure*}

\subsection{Joint analysis with multiple S-stars}\label{sec::discussion:4stars}
\citet{Gillessen2017} has reported the VLT observations of an additional 16 S-stars. 
The combination of multiple S-stars can span a broader range of density profile,
as illustrated by the shaded band in Fig.~\ref{fig::profile}, 
thereby potentially improving the constraint on the spike density profile. 
This subsection is to discuss the outcomes obtained when multiple S-stars are combined.

The VLT data of S1, S9, and S13 are incorporated into the previous S2 dataset,
as suggested by~\citet{Gillessen2017} due to their increased sensitivity in constraining $R_0$ 
and $M_{\rm BH}$.
It can potentially distinguish the extended mass from SMBH, and provide a stringent constraint on the DM spike. 
Including 637 astrometric points and 228 radial velocities, the constraints on the gNFW spike are re-evaluated. 
The $\chi^2$ values of the three stars are combined with Eq.(\ref{eq::chi2_astro_VLT}) and Eq.(\ref{eq::chi2_rv_VLT}), resulting in a total likelihood function containing 39 free parameters.

In Fig.~\ref{fig::4stars_bestfit}, we show the measurements of the four S-stars (points) overlaid with MAP models (solid lines).
The left panel is the celestial coordinates corrected for the offsets and drifts.
The right panel is the radial velocities with respect to the epoch for the stars.
Indeed, we can see that the MAP orbital models can match the measurements reasonably well.

In Fig.~\ref{fig::gnfw_Rsp_4stars}, we present the upper limits on the spike radius $R_{\rm sp}$ with the four S-stars.
The constraints from the combined analysis are slightly stronger than 
the case only including S2 data.
The expected spike radius in the GS model is shown in the black dashed line.
We show that the GS model with $\gamma \geq 0.90$ is excluded by the observations of the four stars at 95\% probability, 
by comparing with $\gamma \geq 0.92$ in the case of S2 only.

In~\citet{GRAVITY2022a}, the GRAVITY data of S2, S29, S38, and S55 are combined to further improve the constraints of extended mass.
We take the publicly available VLT data of S38 and S55~\citep{Gillessen2017} and incorporate them into the previous S2 data set.
In total, we have 348 astrometric points and 122 radial velocities.
Similar joint analyses are performed and the constraints on $R_{\rm sp}$ are presented in Fig.~\ref{fig::gnfw_Rsp_4stars}.
The constraints are slightly stronger than the results using four S-stars, due to the small uncertainties of their semi-major axes and orbital periods.
The GS model with $\gamma \geq 0.83$ is excluded at 95\% probability.

\subsection{Interpretations of GRAVITY present and future S2 sensitivity of the halo extended mass on DM spike}

The GRAVITY coherently combines the light of the Very Large Telescope Interferometer and 
presents $\sim 10~\mu \rm as$-level photometric observations of S-stars~\citep{GRAVITY2017}. 
In~\citet{GRAVITY2022a}, the S2 data collected from 1992.2 to 2021.6 were adopted to constrain the extended mass.
They have found that no more than $2400M_{\odot}$ or $7500M_\odot$ extended mass could exist within 
the apocentre of S2 at $1\sigma$ or $3\sigma$ level respectively.
In addition, \citet{Heissel2022} have estimated a prospect that 
the GRAVITY could gather a full-orbit data of S2 with $50~\rm \mu as$ and $10~\rm km\,s^{-1}$ precision by 2033.
Their result shows that the future S2 sensitivities from GRAVITY could restrict 
the dark extended mass less than $1000M_\odot$ at $1\sigma$ level. 
Hence, in this section we simply project the GRAVITY extended mass limits 
(both present combined limits and the future sensitivities) on $R_{\rm sp}$ and $\gamma_{\rm sp}$.

We learn from the left panel of Fig.~\ref{fig::gnfw_Rsp_s2} that 
the enclosed mass within the apocentre of S2 only weakly depends on the density slope.
We can use the extended mass constraints derived from the Plummer or Bahcall-Wolf cusp profiles in~\citet{GRAVITY2022a} and \citet{Heissel2022} as approximated upper limits of the extended mass of the DM spike, even though the slopes of the two profiles are different from the DM spike.
In this way, we directly convert the extended mass upper limits to the constraints of DM spike parameters.

The left panel of Fig.~\ref{fig::gnfw_gravity} shows the upper limits on the spike radius $R_{\rm sp}$ for the gNFW spike.
The purple solid and dashed lines correspond to the $1\sigma$ and $3\sigma$ upper limts on $R_{\rm sp}$ by 
using the GRAVITY data collected before 2021.6.
For the DM halo with $\gamma=1$, the $1\sigma$ and $3\sigma$ upper limits are 8.0~pc and 19~pc, respectively. 
Our results are in good agreement with the ones obtained by fitting the GRAVITY orbital data ($R_{\rm sp}\lesssim 10~\rm pc$)~\citep{GRAVITY2020a}.
The black dashed line in the figure shows the corresponding $R_{\rm sp}$ in the GS spike model [Eq.~\eqref{eq::gs_spike_radius}].
The gNFW spike with the initial density slope $\gamma$ steeper than $0.64$/$1.00$ is rejected by the data of the current GRAVITY S2 measurements at $1\sigma$/$3\sigma$ level.
The future sensitivities of $R_{\rm sp}$ are presented by the brown solid lines in the case of 
using a full-orbit GRAVITY data of S2.
The $1\sigma$ upper limit on $R_{\rm sp}$ for the gNFW with $\gamma=1$ is $\sim 4~\rm pc$ in this case, 
and the GS spike with $\gamma \gtrsim 0.4$ can be probed.

The right panel of Fig.~\ref{fig::gnfw_gravity} illustrates the limits of the spike slope $\gamma_{\rm sp}$ for the gNFW spike.
Given the GRAVITY measurements collected before 2021.6, 
the $1\sigma$ ($3\sigma$) upper limits on $\gamma_{\rm sp}$ for the gNFW with $\gamma=1$ are 2.21 (2.34).
Once a full-orbit GRAVITY data of S2 were available, 
the prospective $1\sigma$ limit could be improved, and the GS spike with $\gamma_{\rm sp}>2.10$ can be completely probed.
With the GRAVITY S2 orbit only, it is far to detect the DM spike slope $\gamma_{\rm sp} \sim 1.5$,  
corresponding to the DM particles efficiently heated by the dynamical processes.

We also estimate the GRAVITY constraining power for the Einasto spike.
The $1\sigma$ ($3\sigma$) upper limits on $R_{\rm sp}$ for the \einn and \einw benchmarks are 15~pc (30~pc) and 45~pc (83~pc), respectively.
The GS model for the \einn benchmark ($R_{\rm sp}^{\rm GS} = 31.2~\rm pc$) is excluded at $3\sigma$ level.
If adopting the full-orbit GRAVITY measurements, the prospective $1\sigma$ upper limits could be strengthened to 9~pc and 29~pc for \einn and \einw, respectively.
The GS Einasto spike model can be entirely probed by the future full-orbit GRAVITY data of S2.

\subsection{The effect of dark matter annihilation}\label{sec::annihilate}
The DM spike distribution is widely used in the DM indirect detection, 
DM particles could annihilate with each other and produce detectable signals. 
In previous sections, we have discussed the cases of lacking DM annihilation.
Since the DM annihilation can reduce the spike density, as long as the annihilation cross section $\left< \sigma v \right>$ is large enough, 
the DM spike may also survive from the S2 orbit constraints.
In this subsection, we present such a lower limit on $\left< \sigma v \right>$ for the NFW spike.

Considering that the initial velocity distribution of DM particles is isotropic, 
the particles with apicentres outside $R_{\rm sat}$ can also contribute to the density inside $R_{\rm sat}$, 
leading to a weak cusp with a slope of 0.5 for the $s$-wave annihilation~\citep{Vasiliev2007,Shapiro2016}.
The density profile below is adopted~\citep{Vasiliev2007}
\begin{equation}\label{eq::spike_ann2_approx}
    \rho_{\rm ann}(r) =
    \begin{cases}
        0 & \qquad r \leq 2R_{\rm sch},\\
        \rho_{\rm sat}/\sqrt{r/R_{\rm sat}} & \qquad 2R_{\rm sch} < r \leq R_{\rm sat},\\
        \rho_{\rm sp}^{\rm GS}(r) & \qquad R_{\rm sat} < r < R_{\rm sp}^{\rm GS} ,
    \end{cases}
\end{equation}
where $\rho_{\rm sat} \equiv m_{\rm DM}/(\left< \sigma v \right> \tau_{\rm BH})$ is the saturation density,
$R_{\rm sat}$ is the saturation radius defined by $\rho_{\rm sp}^{\rm GS}(R_{\rm sat})=\rho_{\rm sat}$,
$m_{\rm DM}$ is the DM particle mass,
and $\tau_{\rm BH}=10~\rm Gyr$ is the age of the SMBH in the GC~\citep{Lacroix2014}.
The NFW spike density distribution and the spike radius in the GS model are $\rho_{\rm sp}^{\rm GS}(r)$ and $R_{\rm sp}^{\rm GS}$, respectively.

We choose the saturation density $\rho_{\rm sat}$ as the free parameter, set a flat prior for the enclosed mass within the apocentre of S2, and rerun the MCMC sampler using the combined Keck and VLT data.
By marginalizing posterior probability, the 95\% credible upper limit of $\rho_{\rm sat}$ is $\sim 2.2 \times 10^{9}~M_\odot\rm \, pc^{-3}$.
The corresponding saturation radius is $R_{\rm sat} \sim 4.8~\rm mpc$.
Therefore, the surviving NFW spike infers the 95\% lower limit of the annihilation cross section:
\begin{equation}
    \left< \sigma v \right> \gtrsim 7.7\times 10^{-27}~{\rm cm^3\,s^{-1}} \times \left( \frac{m_{\rm DM}}{100~{\rm GeV}} \right) \left( \frac{10~{\rm Gyr}}{\tau_{\rm BH}} \right).
\end{equation}
Interestingly, such a request is indeed satisfied by a good fraction of the parameter space of the inert two Higgs doublet model for the W-boson mass excess \citep{2022PhRvL.129i1802F}.

\section{Summary}\label{sec::summary}
The density of DM can be significantly increased due to the adiabatic growth of BH, creating a structure so-called DM spike.
In this work, we constrain the parameters of the DM spike profile in the GC by using the precise measurements of the S-stars from the Keck and VLT observatories. 
The S-star coordinates and radial velocities are determined by solving the 1PN dynamical equation. 
Utilizing pre-2019 Keck data~\citep{Do:2019txf} and pre-2017 VLT data~\citep{Gillessen2017}, 
we perform a Bayesian analysis with the MCMC algorithm to derive constraints.

We firstly examine the gNFW spike model with the astrometric and spectrometric data of S2 obtained from VLT and Keck telescopes. 
The predicted GS model is rejected with a 95\% credibility if the spike radius $R_{\rm sp}>15.7$~pc or the spike slope $\gamma_{\rm sp}>2.32$, assuming the spike grows from the initial NFW profile ($\gamma=1$).
Our limits of $R_{\rm sp}$ (Fig.~\ref{fig::gnfw_Rsp_s2}) is more stringent than \citet{Lacroix:2018zmg}, because the accurate measurements at the S2 pericentre are used.
We find that the GS spike model with $\gamma \geq 0.92$ is excluded at 95\% level.
If we combine the data from multiple S-stars, only the GS model with $\gamma < 0.83$ can survive the constraints (see Fig.~\ref{fig::gnfw_Rsp_4stars}).

In Sec.~\ref{sec::profile}, by taking the Einasto profile as initial density, 
we derive the GS spike density distribution using the circular-orbit approximation (Fig.~\ref{fig::profile}). 
We find in Fig.~\ref{fig::einasto_index} that the spike profile can be well described by a power-law model 
with a slope of $\sim 2.27$ within the spike radius.
In this work, we simply take two representative Einasto model benchmarks as a initial profile, 
\einn from the simulation and \einw from the {\it Gaia} observation. 
We find that the 95\% upper limits of $R_{\rm sp}$ are 21.5~pc for \einn but 
61.4~pc for \einw (Fig.~\ref{fig::ein_Rsp_s2}). 
Comparing with the theoretical GS spike profile, the \einn case is excluded by the likelihood using the S2 orbit, 
while \einw still survives marginally.

Moreover, we convert the upper limits of the extended mass given by the GRAVITY observation collected before 2021.6 to the halo spike parameters.
As shown in Fig.~\ref{fig::gnfw_gravity}, the $1\sigma$ upper limits on the $R_{\rm sp}$ and $\gamma_{\rm sp}$ are 8.0~pc and 2.21 respectively for the NFW model, and the gNFW spike model with $\gamma \geq 0.64$ is excluded at $1\sigma$-level.
If the simulated full-orbit GRAVITY data of S2 is utilized, 
the GS spike profile with $\gamma \gtrsim 0.40$ can be restricted.
Nevertheless, it can be difficult to detect the DM spike slope $\gamma_{\rm sp} \sim 1.5$ which is a theoretical prediction by including the star distribution or dynamical processes.

Finally, we consider the NFW spike profile with DM annihilation.
If DM particles can annihilate with each other, the density profile in the inner region can be flattened to a weak cusp and thereby may survive the S2 constraints.
Using the S2 orbital data from the two observatories, the 95\% upper limit on the saturation density $\rho_{\rm sat}$ is $\sim 2.2\times 10^9~M_\odot\,\rm pc^{-3}$.
Therefore the cross section needs to be larger than $\sim 7.7\times 10^{-27}~{\rm cm^3\,s^{-1}} ( m_{\rm DM}/100~{\rm GeV})$ at 95\% level.

In the future, more high-resolution data on the S-stars will be collected by the GRAVITY and Keck telescopes.
Several S-stars, such as S21, S23, S24, and S60, will pass their pericentric points in the next few years~\citep{Gillessen2017}. 
In addition, some faint S-stars, such as S62, S4711, S4714, with orbital periods less than 10~yrs or pericentric distances less than several hundreds of $R_{\rm sch}$ are reported by~\citep{Peissker2020b,Peissker2020a,GRAVITY2021b}.
These continuous observations can further probe the DM spike parameters.

\section*{Acknowledgements}
We would like to thank Tuan Do and Stefan Gillessen for sharing us with the data of S-stars.
We appreciate the useful information and helpful discussions from Ran Ding, Lei Feng, Andrew Fowlie, Zi-Qing Xia, Kai-Kai Duan, Zhi-Hui Xu and Bing Sun.
We are grateful to Xiang Li for the spare workstation.
This work is supported by the National Key Research and Development Program of China (No. 2022YFF0503304), the National Natural Science Foundation of China (No. 12003074), and the Entrepreneurship and Innovation Program of Jiangsu Province.

The following software is adopted:
{\tt NumPy} \citep{numpy2020}, {\tt SciPy} \citep{scipy2020}, {\tt Matplotlib} \citep{matplotlib2007}, {\tt Astropy} \citep{astropy2018}, {\tt emcee} \citep{emcee2013}, {\tt iminuit}~\citep{iminuit2020,Minuit1975}.

\section*{data availability}
The astrometric data and radial velocity used in this work are obtained from~\citet{Do:2019txf} and~\citet{Gillessen2017}.
The constraints on the DM spike are presented in Tab.\ref{tab::constraints}.


\appendix

\section{The precession angles of S2 with DM spike}\label{app::precession}

\begin{figure}
    \centering
    \includegraphics[width=\columnwidth]{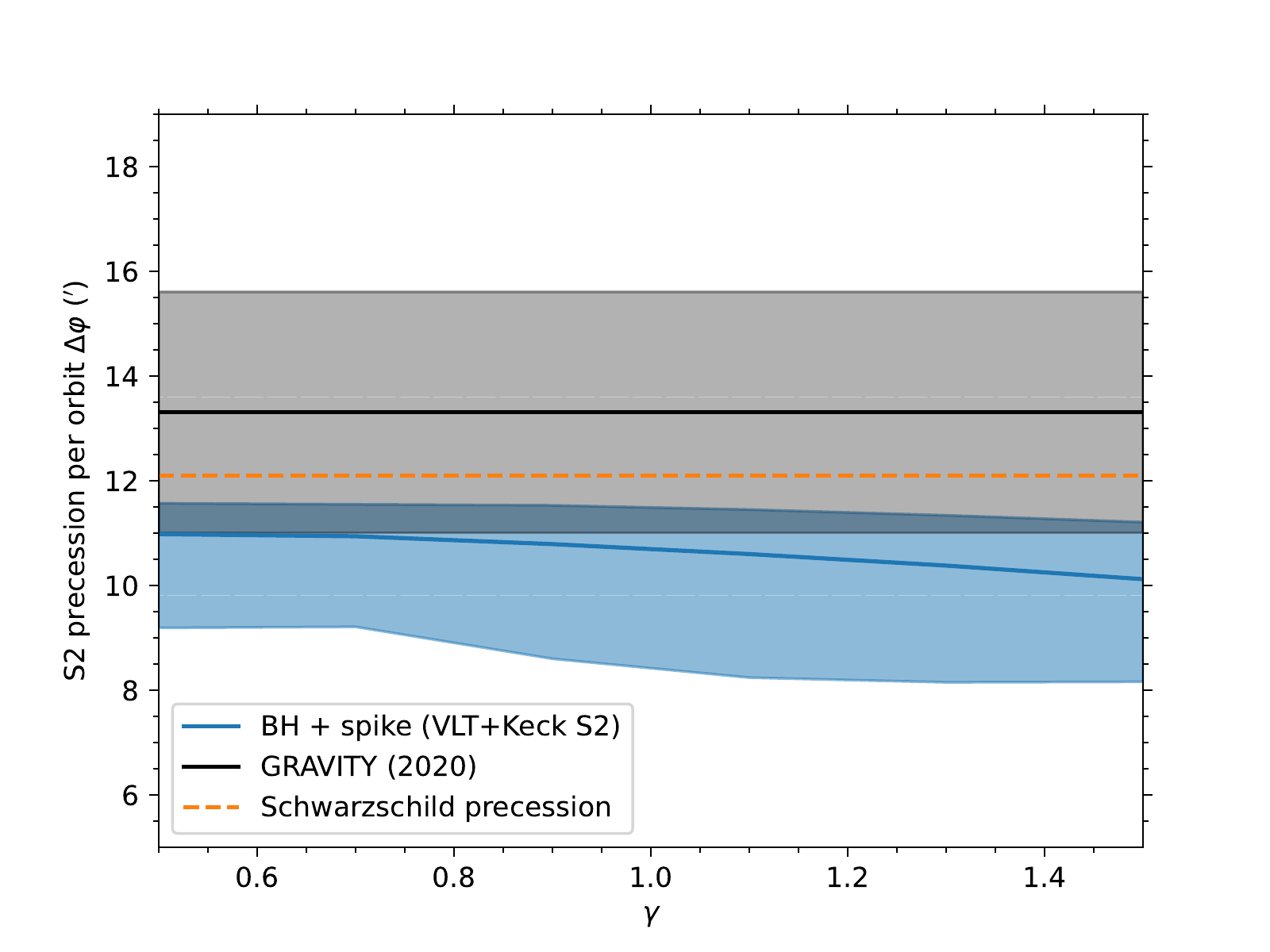}
    \caption{\label{fig::s2prec}
        The S2 precession angle per orbit $\Delta \varphi$ when DM spike with initial NFW slope $\gamma$ exists around the SMBH.
        The blue solid line and blue band correspond to the median and $68\%$ uncertainties of the precession angle respectively.
        The fitted value and the uncertainty measured by the GRAVITY~\citep{GRAVITY2020a} are shown with the black solid line and black band, respectively.
        The orange dashed line illustrates the Schwarzschild precession angle.
    }
\end{figure}
\bigskip

Extended mass such as the DM spike will introduce a retrograde effect in addition to the standard prograde Schwarzschild precession~\citep[e.g.][]{Weinberg2005,Arguelles2022}.
We present in Fig.~\ref{fig::s2prec} the precession angle of S2 per orbit in the case that there is a DM spike with $R_{\rm sp}$ constrained.
The precession angle is calculated as the difference in azimuthal angles between two consecutive pericentre passages, namely, $\Delta \varphi \equiv \varphi(t_{\rm per2})-\varphi(t_{\rm per1})$.
The medians and $1\sigma$ uncertainties for different initial NFW slopes $\gamma$ are shown with the blue solid line and blue band, respectively.
The median precession angles in our model are $10'-11'$ per orbit, slightly smaller than the Schwarzschild precession angle of $\approx 12'$ (the orange dashed line) due to the retrograde effect.
We also show the parameterized Schwarzchild precession angle of $(1.10\pm0.19)\times 12.1'$ measured by the GRAVITY~\citep{GRAVITY2020a} with the grey band.
Our results are consistent with the measurement within $\sim 1\sigma$ statistical uncertainty.

\section{The constraints on the generalized NFW spike}
\begin{table}
    \centering
    \caption{
        The 95\% credible upper limits on the spike radius $R_{\rm sp}$ or spike slope $\gamma_{\rm sp}$ for different initial NFW density slope $\gamma$.
        The values in the second and fifth columns are the values predicted by~\citet{Gondolo:1999ef}.
        The third and sixth columns show the constraints using the VLT and Keck data of S2.
        The fourth column presents the constraints using the combined observations of 3 S-stars.
        See Sec.~\ref{sec::results} and Sec.~\ref{sec::discussion:4stars} for detail.
    }\label{tab::constraints}
    \begin{tabular}{lccc|cc}
     \hline
     $\gamma$ & $R_{\rm sp}^{\rm GS}$ & $R_{\rm sp,95\%}^{\rm S2}$ & $R_{\rm sp,95\%}^{\rm 3stars}$ & $\gamma_{\rm sp}^{\rm GS}$ & $\gamma_{\rm sp,95\%}^{\rm S2}$ \\
              &      (pc)       &      (pc)            & (pc) & & \\
     \hline
     0.5 & 37.9 & 71.9 & 55.5 & 2.285 & 2.45 \\
     0.6 & 34.3 & 55.4 & 45.2 & 2.294 & 2.42 \\
     0.7 & 30.3 & 44.9 & 41.2 & 2.303 & 2.37 \\
     0.8 & 26.3 & 32.5 & 27.6 & 2.313 & 2.36 \\
     0.9 & 22.5 & 23.5 & 19.6 & 2.323 & 2.33 \\
     1.0 & 18.6 & 15.7 & 12.5 & 2.333 & 2.32 \\
     1.1 & 14.8 & 10.2 &  9.0 & 2.345 & 2.28 \\
     1.2 & 11.4 &  5.6 &  5.2 & 2.357 & 2.25 \\
     1.3 & 8.42 &  3.0 &  3.0 & 2.370 & 2.23 \\
     1.4 & 5.95 &  1.4 &  1.4 & 2.385 & 2.18 \\
     1.5 & 3.96 & 0.56 & 0.53 & 2.400 & 2.14 \\
     \hline
    \end{tabular}
\end{table}

The 95\% credible constraints on the spike radius $R_{\rm sp}$ or spike slope $\gamma_{\rm sp}$ in Fig.~\ref{fig::gnfw_Rsp_s2}, Fig.~\ref{fig::gnfw_Rsp_4stars} and Fig.~\ref{fig::gnfw_gammasp} are shown in Tab.\ref{tab::constraints}.

\bsp	
\label{lastpage}
\end{document}